\documentstyle[11pt,epsfig]{article}

\newcommand{\arcsecd}{\mbox{.\kern -3pt{${}^{\prime\prime }$}}}
\newcommand{\micron}{$\mu{\rm m}$}
\newcommand{\ch}{${\rm CH_4}$}
\newcommand{\nh}{${\rm NH_3}$}
\newcommand{\nhsh}{${\rm NH_4SH}$}
\newcommand{\ph}{${\rm PH_3}$}
\newcommand{\chd}{${\rm CH_3 D}$}
\newcommand{\geh}{${\rm GeH_4}$}
\newcommand{\ash}{${\rm AsH_3}$}
\newcommand{\ho}{${\rm H_2 O}$}
\newcommand{\h}{${\rm H_2}$}
\newcommand{\cm}{${\rm cm^{-1}}$}
\newcommand{\degre}{$^{\circ}$}
\newcommand{\ratio}{${\rm ^{15}N/^{14}N}$}
\newcommand{\inh}{${\rm ^{15}NH_3}$}
\newcommand{\rnh}{${\rm ^{14}NH_3}$}
\begin{document}

\begin{center}
{\Large\bf ISO-SWS observations of Jupiter: measurement of the ammonia tropospheric profile and of the $^{15}$N/$^{14}$N isotopic ratio}\\[1.5cm]

Thierry Fouchet$^1$, Emmanuel Lellouch$^1$, Bruno B\'ezard$^1$, Th\'er\`ese Encrenaz$^1$, Pierre Drossart$^1$, Helmut Feuchtgruber$^2$ and Thijs de Graauw$^3$\\[1.5cm]

$^1$DESPA, Observatoire de Paris, 5 Place Jules Janssen, 92195 Meudon Cedex, France\\
tel.: 33.1.45.07.76.60\\
fax : 33.1.45.07.71.10\\
email : Thierry.Fouchet@obspm.fr\\[0.5cm]

$^2$Max-Planck-Institut f\"ur Extraterrestrische Physik, 85748 Garching, Germany\\[0.5cm]

$^3$SRON, PO box 800, 9700 AV Groningen, The Netherlands\\[1.5cm]

57 pages\\
2 tables\\
15 figures\\
\end{center}
\newpage

\noindent Proposed running title: Ammonia in Jupiter from ISO-SWS observations\\

\noindent Send correspondence to: Thierry Fouchet, Observatoire de Paris, DESPA-T1m, 5 Place Jules Janssen, 92195 Meudon Cedex, France; tel.: 33.1.45.07.76.60; fax: 33.1.45.07.71.10; email: Thierry.Fouchet@obspm.fr
\newpage

\section*{Abstract}

\indent\indent We present the results of the Infrared Space Observatory Short Wavelength Spectrometer (ISO-SWS) observations of Jupiter related to ammonia. We focus on two spectral regions; the first one (the 10-\micron\ region), ranging from 9.5 to 11.5\,\micron, probes atmospheric levels between 1 and 0.2\,bar, while the second one (the 5-\micron\ window), ranging from 4.8 to 5.5\,\micron, sounds the atmosphere between 8 and 2\,bars. The two spectral windows cannot be fitted with the same ammonia vertical distribution. From the 10-\micron\ region we infer an ammonia distribution of about half the saturation profile above the 1-bar level, where the N/H ratio is roughly solar. A totally different picture is derived from the 5-\micron\ window, where we determine an upper limit of $3.7\times 10^{-5}$ at 1\,bar and find an increasing \nh\ abundance at least down to 4\,bar. This profile is similar to the one measured by the Galileo probe (Folkner {\em et al.}\ 1998). The discrepancy between the two spectral regions most likely arises from the spatial heterogeneity of Jupiter, the 5-\micron\ window sounding dry areas unveiled by a locally thin cloud cover (the 5-\micron\ hot spots), and the 10-\micron\ region probing the mean jovian atmosphere above 1 bar. The \inh\ mixing ratio is measured around 400\,mbar from $\nu_2$ band absorptions in the 10-\micron\ region. We find the atmosphere of Jupiter highly depleted in ${\rm^{15}N}$ at this pressure level (${\rm (^{15}N/^{14}N)_{J}=(1.9^{+0.9}_{-1.0})\times10^{-3}}$, while ${\rm (^{15}N/^{14}N)_{\oplus}=3.68\times 10^{-3}}$). It is not clear whether this depletion reveals the global jovian \ratio\ ratio. Instead an isotopic fractionation process, taking place during the ammonia cloud condensation, is indicated as a possible mechanism. A fractionation coefficient $\alpha$ higher than 1.08 would explain the observed isotopic ratio, but the lack of laboratory data does not allow us to decide unambiguously on the origin of the observed low \ratio\ ratio.\\

{\em Key words:} Atmospheres, Composition; Infrared Observations; Jupiter, Atmosphere
\newpage

\section{Introduction}

\indent\indent Ammonia plays an important role in understanding the jovian atmosphere in three main areas. First, its bulk abundance is an indicator of the source of planetesimals which have enriched Jupiter's atmosphere in C, S, N and presumably O (${\rm C/H=2.9\times solar}$, ${\rm S/H=2.5\times solar}$ (Niemann {\em et al.}\ 1998)). As discussed by Owen {\em et al.}\ (1997) a jovian C/N ratio larger than solar would indicate that such planetesimals formed at too warm a temperature to trap N$_2$ in their ices, as it is the case for the comets (Eberhardt 1998). This picture is consistent with the deep \nh\ mixing ratio deduced by Bjoraker {\em et al.}\ (1986), de Pater (1986) and Joiner and Steffes (1991), approximately $2.5\times10^{-4}$ (${\rm1.1\times solar}$, assuming the solar N/H of Anders and Grevesse (1989)), but is challenged by Lellouch {\em et al.}\ (1989) and Carlson {\em et al.}\ (1993) who inferred a deep value of $4.5\times10^{-4}$ (${\rm2.0\times solar}$), and more recently by the Galileo probe which measured ${\rm(4\pm0.5)\times solar}$ at 8\,bars (Folkner {\em et al.}\ 1998).

Secondly, the \ratio\ isotopic ratio in the jovian atmosphere is thought to indicate the primitive \ratio\ ratio in the pre-solar nebula. Since the measurement of the \ratio\ ratio in the solar wind exhibits large variations, ($[-21\%,+12\%]$), with respect to the terrestrial \ratio\ (Geiss and Bochsler, 1982), Jupiter is the best suited object to determine this important parameter for models of the solar system formation. However, an accurate determination of the \ratio\ ratio from the 10-\micron\ region has proved to be difficult. Encrenaz {\em et al.}\ (1978) and Tokunaga {\em et al.}\ (1980) deduced a value in agreement with the terrestrial ratio, but with uncertainties of a factor of 2.

Finally, as a condensable species, ammonia is a major tracer of the dynamical and meteorological state of the planet. Indeed, at several bars, ammonia may first dissolve in liquid water, then it reacts with H$_2$S to form an \nhsh\ cloud at 2 bars, condenses to form the \nh-ice cloud around 0.6\,bar and is finally photodissociated in the upper troposphere. In agreement with this model, the gaseous ammonia mixing ratio is found to sharply decrease with altitude above the 0.6-bar pressure level (Kunde {\em et al.}\ 1982; de Pater 1986), while analysis of the Voyager/IRIS spectra (Carlson {\em et al.}\ 1993) and of ground based centimetric observations (de Pater, 1986) suggested a constant mixing abundance at pressures deeper than 2\,bars. However, new insights came from the Galileo probe, since the ammonia mixing ratio at entry point was observed to increase down to 8 bars and to remain roughly constant at deeper levels (Folkner {\em et al.}\ 1998).

The uncertainties and the discrepancies between the above studies clearly show that our knowledge of ammonia has not yet settled on a unique picture. The Short Wavelength Spectrometer on board the Infrared Space Observatory (ISO-SWS) offered a unique opportunity for a simultaneous coverage of Jupiter's spectrum from 2.4 to 16\,\micron\ at an average resolution of 1200. Here we present the analysis of the ISO-SWS data regarding ammonia. We use two spectral regions, the 5-\micron\ window and the 10-\micron\ region, to infer the ammonia vertical distribution from 8 to 0.2\,bar. In addition, the spectral resolution of the measurements is sufficient to determine the \ratio\ isotopic ratio. The observations are presented in Section~2. Section~3 describes the atmospheric model used to analyze the spectrum. From the comparison of ISO-SWS data and synthetic spectra, we infer the \nh\ vertical profile and the \ratio\ ratio. Finally, the results are discussed in Section~4.

\section{Observations}

\indent\indent Descriptions of the ISO satellite and of the SWS can be found respectively in Kessler {\em et al.}\ (1996) and de Graauw {\em et al.}\ (1996). ISO-SWS grating observations of Jupiter were obtained on December 6th, 1996 and May 23rd, 1997 using respectively the AOT 06 and AOT 01 modes of SWS. The AOT 06 observations are at a slightly higher resolution, but analysis of the two spectra leads to the same conclusions. Therefore this paper will be based on the interpretation of the AOT 01 spectrum. The entire spectral region from 2.4 to 45\,\micron\ was observed, but saturation effects due to the very high flux of Jupiter limited the useful range to 2.4--16\,\micron. The exposure time was 110 minutes and the resolving power ranges from 800 to 1500. The aperture, ${\rm14\times20\,arcsec^2}$ at $\lambda<12.5\,$\micron\ and ${\rm14\times27\,arcsec^2}$ at $\lambda>12.5\,$\micron, was centered on the planet with the long axis aligned perpendicular to the ecliptic, thus roughly parallel to the N-S polar axis. The sub-earth latitude was 0.07\degre, the field of view thus covering latitudes between $-30$\degre\ and 30\degre. The average airmass over the regions of jovian disk encompassed by the aperture is 1.08. The SWS flux and wavelength calibration are described in Schaeidt {\em et al.}\ (1996) and Valentijn {\em et al.}\ (1996). The accuracy of the absolute flux scale is approximately 20\%.

\section{Radiative transfer analysis}

\indent\indent We have interpreted the data with a line-by-line radiative transfer code. Our model included the rovibrational bands from \ch, \nh, \ph, \geh, \ho, \ash\ and \chd. The parameters used in the modeling for these different gases ---i.e.\ the mole fractions, the line positions, the strengths and lower energies, and the \h\ and He broadening coefficients--- are listed in Table~I. The line profiles were assumed to be Voigt-shaped (i.e.\ essentially Lorentz-shaped at the sounded pressures). To account for the expected sub-lorentzian behavior of the far wings, we applied a cutoff at 20\,\cm. The collision-induced absorption of \h--\h\ and \h--He is also an important source of opacity. It was included in our calculation, using the theoretical works of Borysow {\em et al.}\ (1985, 1988). Cloud layers, needed to fit the observed continuum level, were introduced with adjustable pressure levels, transmissions and scale heights, but our radiative transfer code does not explicitly take into account scattering effects. %{\bf
However, we have estimated the modifications to the retrieved gaseous abundances induced by multiple scattering, as described in Section 3.1.2.
%}

The 5-\micron\ and 10-\micron\ measurements sound Jupiter's troposphere. Therefore, a tropospheric temperature profile is needed for interpreting the observations. It is usually retrieved using the S(0) and S(1) \h-lines (Gautier {\it et al.}, 1981). However, because of the saturation of the ISO-SWS spectrum longward of 16\,\micron, we used the temperature profile measured in-situ by the Galileo probe (Seiff {\em et al.}\ 1998), which we slightly modified in the upper troposphere. Indeed, the Galileo probe entered a peculiar region of Jupiter's atmosphere ---the edge of a 5-\micron\ hot spot--- which is not representative of the large area covered by the ISO-SWS aperture. This is consistent with the fact that the brightness temperature of 115.7\,K, as calculated in the \h--He continuum at 602\,\cm\ for the Galileo temperature profile, is slightly colder than the average 117.5\,K observed by Voyager/IRIS at this wavenumber, between $-30$\degre\ and $+30$\degre\ latitude (Hanel {\em et al.}\ 1979). %{\bf
To account for this, the Galileo temperature profile was smoothed over one scale height to erase local oscillations, then warmed by 2\,K at 150\,mbar
%}
and smoothly reconnected to the original Galileo profile at 30 and 500\,mbar. In the lower troposphere, below 500 mbar, the temperature field is expected to be homogenized by convection, and we adopted there the Galileo probe measurements. While spatial variations do occur in the lower stratosphere, the choice of the temperature profile in that region is not critical for the results for ammonia, because of the strong depletion due to \nh\ condensation and photodissociation in this region.

\subsection{${\rm\bf NH_3}$ vertical distribution}

\subsubsection{Upper troposphere: 10-${\rm\bf\mu m}$ region}

\indent\indent The 10-\micron\ region is dominated by the absorptions of the $\nu_2$ band of ammonia, leaving the retrieval of the ammonia vertical profile free of contamination from other gases. Figure~1 displays the contribution functions in the Q-branch center at 10.73\,\micron\ and in a line center at 10.83\,\micron, as well as in the pseudo-continuum at 10.91\,\micron\ between \nh\ lines. The widths of the contribution functions show that three independent points on the \nh\ vertical profile can be inferred at 300, 410 and 630\,mbar. At pressures less than 300\,mbar and larger than 630\,mbar the \nh\ mixing ratio was extrapolated with the same slope as the one between, respectively, 300 and 410\,mbar, and 410 and 630\,mbar. At pressures less than 200\,mbar the ammonia mixing ratio has only a marginal influence on the calculated spectra. As \nh\ is removed by photodissociation in the upper troposphere, we cut it off at this pressure level. Below 900\,mbar, following our results of Section 3.1.2, the distribution was held vertically constant as soon as it reached the value of ${\rm 3.5\times 10^{-4}}$ (${\rm 1.6\times solar}$).

The cloud layer around 700\,mbar also contributes to the atmospheric opacity, making the retrieved ammonia abundance at 630\,mbar sensitive to the cloud transmission. Therefore we included this layer in our model. %{\bf
We assumed that the cloud layer is composed of \nh-ice particles, as expected from thermochemical models and indicated by the ISO-SWS spectrum at 2.9\,\micron\ (Brooke {\it et al.}, 1998). The cloud transmission was spectrally calculated as ${\rm T_c} = \exp(-K.Q_{abs})$ where $Q_{abs}$, the particle absorption cross section, was calculated from Mie theory for \nh-ice spheres, using optical constants measured by Schmitt {\it et al.}\ (1998). The factor $K$ was adjusted to obtain the correct flux at 11.08\,\micron.  Following Brooke {\em et al.}\ (1998), we first used a particle radius of 10\,\micron. Following Conrath and Gierasch (1986), we used a particle scale height (${\rm H_p}$) equal to 0.15 times the atmospheric scale height (${\rm H_p = 0.15\times H_g}$). The result of the synthetic calculation is compared with the ISO-SWS spectrum in Fig.~2a (dashed line). Clearly, ammonia ice particles of this size give too much absorption in the ammonia ice band centered at 9.5\,\micron (see mismatches at 9.62 and 9.82\,\micron). Our crude modeling may not exactly estimate the transmission of a scattering ammonia cloud, but similar results were obtained by Marten {\it et al.}\ (1981) with a multiple scattering code. They essentially concluded that the absence of any ammonia ice feature in the Voyager/IRIS spectra implies that the cloud particles at 700\,mbar act as grey scatterers. Since the ISO-SWS spectrum is also exempt of such a feature, we reach the same conclusion. Grey scatterers can be obtained for \nh-ice particle larger than 100\,\micron\ (see Fig.~2b). Reasons for the disagreement with Brooke {\it et al.}\ (1998) will be investigated in a future publication. In the rest of this paper, we consider the cloud particles to be grey absorbers at 10\,\micron.
%}

The cloud transmission was set to the value deduced from Voyager/IRIS observations in the Equatorial Zone by Gierasch {\em et al.}\ (1986), i.e.\ $\rm{T_c = 0.8}$. The results of Gierasch {\em et al.}\ (1986) are found to be compatible with the cloud structure measured by Galileo/SSI (Banfield {\em et al.}\ 1998); since the Voyager/IRIS results are at a larger spatial scale we used them here. The influence of this parameter will be fully explored in Section 3.2. We used a particle scale height (${\rm H_p}$) equal to 0.15 times the atmospheric scale height (${\rm H_p = 0.15\times H_g}$).

The best fit between the observed and synthetic spectra is shown in Fig.~2a (dotted line) for the nominal temperature profile and the atmospheric parameters mentioned above. The corresponding \nh\ profile is shown in Fig.~3 (solid line). While a general match of the observed spectrum is achieved, departures can be seen between the ISO-SWS and synthetic spectra, in the Q-branches and P and R-branches. The flux level in the core of the Q-branches (notably at 10.73\,\micron) is overestimated by the model, while that in the core of most of the P and R-lines is underestimated (notably at 10.08, 11.01 and 11.27\,\micron). In addition, the synthetic spectrum is too low in the wings of the Q-branches (10.20 and 10.90\,\micron), and too high in the continuum between the P and R lines (10.0 and 11.1\,\micron\ for example). These systematic discrepancies were already observed by previous authors (Griffith {\em et al.}\ 1992) and their origin remains unclear. They could be due to a limitation of spectroscopic data or to uncertainties in the real lineshape of the far wings which could possibly significantly differ from the assumed sub-lorentzian profile. This error limits the accuracy of our retrieved \nh\ abundance to  $_{-45\%}^{+36\%}$ at 300\,mbar and $_{-32\%}^{+50\%}$ at 630\,mbar. Other mismatches between the synthetic and observed spectra occur at 9.60, 9.79, 9.97, 10.60, 11.07 and 11.31\,\micron. They are related to the \ratio\ isotopic ratio, and will be fully analyzed in Section 3.2.

Uncertainties in the temperature profile also induce important uncertainties on the vertical ammonia distribution. We have allowed the temperature profile to vary between ${\rm\pm2\,K}$ of the nominal temperature profile. This variation is derived from the Voyager/IRIS brightness temperature map of Jupiter at 602\,\cm\ (Hanel {\em et al.}\ 1979), where departures of ${\rm\pm2\,K}$ of the brightness temperature are observed around a mean value of 117.5\,K. Such a variation is certainly conservative for pressure levels deeper than 500\,mbar. This uncertainty in the temperature profile induces an error of a factor of 2--3 on the \nh\ profile (Fig.~3). However, qualitatively, the ammonia profile remains unchanged. It always exhibits a subsaturated region between ${\rm\sim1\,bar}$ and 400\,mbar, and then decreases rapidly with altitude at pressures less than 400\,mbar, with a \nh\ mixing ratio scale height of ${\rm H_q=0.085\times H_g}$.
 
\subsubsection{Lower troposphere: 5-${\rm\bf\mu m}$ window}

\indent\indent The 5-\micron\ window allows the retrieval of the ammonia vertical profile between 2 and 7\,bar. However, overlapping absorptions of \ho, \ph, and of the entire cloud system make this determination more complicated than in the 10-\micron\ region. This is illustrated in Fig.~4, where the respective contributions of \nh\ and \ho\ are displayed. It can be seen that the \ho\ absorptions dominate at $\lambda<5.2$\,\micron, whereas \nh\ absorptions prevail at $\lambda>5.2$\,\micron. The prominent lines at 4.84, 4.90, 4.96, 5.02, 5.08, 5.15 and 5.21\,\micron\ (Fig.~4a) are water absorptions. They enable us to determine two points (Fig.~4b) on the water vertical distribution at 5.9\,bar (4.84\,\micron) and 3.8\,bar (5.21\,\micron), although the 5.9-bars point suffers from some limited contamination by other gases. Absorption of ammonia dominates from 5.3 to 5.5\,\micron, and exhibits two strong spectral features at 5.16 and 5.20\,\micron\ (Fig.~4c). The contribution functions (Fig.~4d) show that these absorptions probe the \nh\ vertical profile down to 4\,bar. At wavelengths shortward of 5\,\micron\ the flux originates from even deeper levels, but in this region the \nh\ lines are strongly blended with the \ho\ and \ph\ lines. Large uncertainties will thus result in the determination of the deep ammonia mixing ratio.

Direct retrieval of the cloud properties from the observed spectrum is very difficult in the 5-\micron\ window, since the cloud particles act as grey absorbers and scatterers (Carlson {\em et al.}\ 1994). Previous authors (B\'ezard {\em et al.}\ 1983; Bjoraker {\em et al.}\ 1986; Lellouch {\em et al.}\ 1989; Roos-Serote {\em et al.}\ 1998) found that the effect of clouds can be modeled phenomenologically by using a purely transmitting cloud in the 1.5--2.0 bar range. We adopted this approach with the following modeling strategy: we assumed a cloud model (i.e.\ a cloud base pressure and scale height); for this model, we determined, through an iterative process, the cloud transmission and the vertical distributions of both water and ammonia. The overall quality of the associated fit in turn validated our assumptions on the cloud model. %{\bf
In addition, reflection of solar radiation on the \nh\ cloud was accounted for by introducing a low albedo (0.03) reflecting cloud at 0.55\,bar. This albedo was found necessary to fit the 4.3-\micron\ region (Encrenaz {\em et al.}\ 1996).

Scattering models indicate that optical thickness for the putative \nhsh\ and \ho\ cloud layers may be large (Carlson {\em et al.}, 1994). In this case, neglecting multiple scattering may be a poor approximation. However, we show below that accounting for multiple scattering would not change qualitatively our results.
%}\\

\noindent{\em Spatially homogeneous model}\\

We first calculated the synthetic spectrum for an infinitely thin (${\rm H_p=0}$) cloud located at 1.5\,bar (${\rm\sim190\,K}$). This cloud model is meant to represent planet-average conditions. The water vertical distribution inferred in this case corresponds to 1.5\%
of saturation, a result similar to that obtained by Roos-Serote {\em et al.}\ (1998) who analyzed Galileo/NIMS hot spot observations. Consistent with the Galileo probe Nephelometer measurements (Ragent {\em et al.}\ 1998) and the analysis of Roos-Serote {\em et al.}\ (1998), we do not need extra opacity at deeper levels as would be expected from the condensation of water at the 5-bar pressure level. The ammonia vertical distribution is constrained to fit the oberved line depths at 5.32\,\micron\ (2\,bar), 5.20\,\micron\ (4.4\,bar), and 4.94\,\micron\ (7.3\,bar), and the cloud transmission needed to fit the continuum level is very low (8\%)
(Fig.~5, dotted line).
Strong disagreements between the observed and the synthetic spectra are evident longward of 5.25\,\micron, where the predicted flux level is far too strong. We tried to reduce this disagreement by increasing the ammonia mixing ratio, but it strongly increased the \nh\ line depths shortward of 5.20\,\micron\ as shown on Fig.~5 (dashed line). It also could not be lowered by varying the water mixing ratio, since water is too weak an absorber in this spectral region. The problem is in fact due to the cloud emission which, with an effective brightness temperature of 188\,K, accounts for nearly half of the synthetic flux at wavelenghts greater than 5.25\,\micron. The data rather indicate that the cloud has an effective brightness temperature lower than 166\,K. This can be obtained by different means: raising the altitude of the cloud base, or using a finite cloud scale height, or any combination of the two. Thus, many cloud models are compatible with the ISO-SWS observations of the 5-\micron\ window---from a vertically thin but optically thick layer located at the 1-bar pressure level to a very diffuse cloud located deep in the troposphere. Theoretical models of the cloud chemistry (Atreya 1986; Carlson {\em et al.}\ 1987) predict the condensation of a \nhsh\ cloud with a cloud base at 1.9\,bar, but such a cloud has not been unambiguously detected either by the Galileo probe or by remote sensing. Therefore, we limited our analysis to two extreme cases: a model with an infinitely thin cloud layer located at 1\,bar, and a cloud layer at 1.9\,bar with ${\rm H_p=0.6\times H_g}$. In both cases the cloud transmission is ${\rm T_c=8\%}$ (${\rm\tau_c=2.5}$).

We then attempted to fit the observed spectrum with a vertically uniform ammonia profile. The results of the calculation for a mixing ratio of $1.0\times10^{-4}$ are displayed on Fig.~6 and 7 for a cloud base at 1\,bar and 1.9\,bar, respectively. The \nh\ line at 4.94\,\micron\ is not reproduced deep enough by the model, due to a too low ammonia abundance around 7\,bar, while on the contrary the modeled 5.32-\micron\ line is too strong, the \nh\ mixing ratio being too high at the 2-bar pressure level. A vertically uniform ammonia profile below the condensation level must thus be rejected in favor of a profile decreasing with altitude. Our best fits are displayed on Fig.~6 and Fig.~7, and the inferred ammonia vertical profile on Fig.~8. In this profile the \nh\ mixing ratio is ${\rm [NH_3]/[H_2]=2.8\times10^{-4}}$ (${\rm N/H=1.3\times solar}$) at pressures greater than 4\,bars. However, we will show further in this section that this value suffers from large uncertainties. Above the 4-bar level the \nh\ mixing ratio decreases with altitude, leading to a value of $3.7\times 10^{-5}$ at the 2-bar level.

%{\bf
The retrieved gaseous abundances may be significantly different from the true values, since we have neglected scattering within the cloud layer. However, accounting for multiple scattering is not likely to change our main conclusion of an ammonia vertical distribution increasing with depth. Indeed, light scattering has two major effects on the flux transmitted through the cloud layer. First, the photons are angularly redistributed. This redistribution affects the average flux level of spectrum, but does not change significantly the modeled line-to-continuum ratios because most of the flux originates from atmospheric levels located below the cloud level, and because the extinction cross sections of ice particles are nearly constant for the different cloud types over the whole 5-\micron\ region (see Fig.~7 of Carlson {\em et al.}\ (1994)). Thus, the outgoing flux is equally affected at each wavelength by the angular redistribution, and line-to-continuum ratios are left unchanged.

A second, more important, effect of multiple scattering is to increase the geometrical path of photons within the cloud layer. Therefore, a model that does not account for this effect will tend to overestimate the amount of the molecular absorbers. However, the effect on the retrieved mixing ratios is much more smaller for lines probing below the cloud layer, such as the weak 5.20-\micron\ line, than for lines probing cloud levels, as the strong 5.32-\micron\ line. Thus, accounting for multiple scattering would only strengthen our conclusion of a larger \nh\ mixing ratio at 4\,bars than at 2\,bars.    

We quantitatively estimated this effect by comparison with the results of Carlson {\em et al.}\ (1993) (their Fig.~2).  Carlson {\em et al.}\ found that the effect of multiple scattering in the deep cloud layer (\ho\ cloud at 4.9\,bars) is to decrease the continuum flux at 1980\,\cm\ by 15\%. Using their molecular distributions, we simulated the increase of optical depth due to scattering by multiplying gaseous optical depths in each layer located within the cloud by a factor proportional to the water cloud opacity. The constant of proportionality was adjusted to reproduce the 15\% decrease at 1980\,\cm. As seen in Fig.~9, this empirical approach is able to reproduce the major effects obtained by Carlson {\em et al.}\ (1993): the continuum at 2000--2300 \,\cm is more affected than at 1800--2000\,\cm; the flux in the water and ammonia lines is only slightly reduced by multiple scattering. 

We then applied this approach to the ISO-SWS data, keeping the same relation between the increase of optical thickness and the cloud opacity. To recover a good fit of the data, we had to reduce the \nh\ mixing ratio from their initial values by a few tens of percents at 2\,bars and only by a few percents at 4\,bars. As this approach remains empirical, and for internal consistency, we do not regard these modified mixing ratios as definitive estimates. But we strongly feel that a vertically inhomogeneous \nh\ profile is supported by the data.
%}

The 5-\micron\ observations do not constrain the ammonia abundance at pressures less than 2\,bars, but a vertically uniform extrapolation of the retrieved mixing ratio at 2\,bar gives an upper limit of ${\rm 3.7\times 10^{-5}}$ at 1\,bar. This is in striking contradiction with our analysis of the 10-\micron\ region (Fig.~3), where the \nh\ distribution was found to reach a solar abundance around 1\,bar. We think that this discrepancy highlights the spatial heterogeneity of Jupiter's atmosphere.

Indeed, in the above modeling, the region seen by the SWS aperture (${\rm14\times20\,arcsec^2}$) was assumed to be homogeneous, although this region, from 30\degre\ N to 30\degre\ S, encompasses the NTrZ (North Tropical Zone), the NEB (North Equatorial Belt), the EqZ (Equatorial Zone), the SEB (South Equatorial Belt) and the STrZ (South Tropical Zone). Differences between belts and zones have long been recognized, especially at 5\,\micron, where the NEB exhibits very large spatial contrasts, with notable bright regions, the so-called 5-\micron\ hot spots (Terrile {\em et al.}\ 1979), dominating the overall 5-\micron\ flux. The high flux in the hot spots is attributed to a low opacity of the cloud cover. Carlson {\em et al.}\ (1994) calculated the cloud optical depth to vary from 0.04 in the NEB hot spots to 6.2 in the NTrZ and 10 in the EqZ, making the NTrZ and the EqZ nearly dark at 5\,\micron. Comparison of Galileo/NIMS observations obtained on the day and night sides of Jupiter also concluded to a dark EqZ at 5\,\micron\ (Drossart {\em et al.}\ 1998). Our preferred \nh\ vertical distribution from the 5-\micron\ data is therefore expected to be characteristic of the hot spot regions of Jupiter, while the one derived from the 10-\micron\ spectral range should represent the average distribution of \nh\ at Jupiter's low latitudes.\\

\noindent {\em Spatially heterogeneous model}\\

To validate this hypothesis we constructed a model of Jupiter's atmosphere which includes two distinct regions. The first one, meant to represent the cloud-covered part of Jupiter, has a cloud optical depth of 6.2 (transmission of 0.2\%).
This cloud optical depth, arbitrarily chosen to model an optically very thick cloud cover, cannot be directly compared with the value of Carlson {\em et al.}\ (1994) who accounted for scattering within the cloud. The cloud effective brightness temperature must still be limited to 166\,K, which, as stated above, can be obtained with various cloud models. Possible solutions are a infinitely thin cloud at 1.0\,bar or a cloud base at 1.9\,bar and a vertical scale height ${\rm H_p=0.40\times H_g}$. To be consistent with our analysis of Section 3.1.1 we use, for this region, the ammonia vertical distribution derived from the ISO-SWS observations at 10\,\micron\ (Fig.~3). The second region, meant to represent the hot spots, is cloud-free. Its surface covering factor (6.5\%)
in the SWS aperture and its ammonia and water vertical distributions are adjusted to fit the 5-\micron\ observations. This surface covering factor is lower than the needed cloud transmission in the previous spatially homogeneous model (${\rm T_c=8\%}$, ${\rm\tau_c=2.5}$), since the effective cloud transmission must be calculated for an average airmass of 1.08, decreasing the cloud transmission to an effective value of ${\rm T_{c\,eff}=6.5\%}$ (${\rm\tau_{c\,eff}=1.08\times\tau_c=2.7 \Rightarrow T_{c\,eff}=6.5\%}$). The result of this modeling is displayed on Fig.~10.

The derived ammonia profile in this second region (Fig.~8, dotted line) is similar to the profile determined in the spatially homogeneous model. The ammonia mixing ratio is constant (${\rm[NH_3]/[H_2]=3.5\times10^{-4}}$, ${\rm N/H=1.6\times solar}$) at pressures greater than 4 bars, then it decreases with altitude at least up to the 2-bar level where the retrieved ammonia abundance is $2.7\times 10^{-5}$. Due to the very low transmission in the cloudy region, nearly all the flux emerges from the hot spots, except in the strongest lines of \nh\ at 5.32, 5.42, 5.46 and 5.49\,\micron, leaving the synthetic spectrum almost unchanged by Jupiter's heterogeneity. Disagreements between the observed and synthetic spectra exist at some specific wavelengths (4.86, 5.07 and 5.13\,\micron). They may result from uncertainties on the strengths of some few \nh\ and \ph\ lines, and consequently they notably affect the uncertainties on the retrieved ammonia mixing ratio at pressures greater than 4\,bars, a problem discussed below.

%{\bf
This crude modeling in terms of two regions obviously does not reflect the continuous changes of the jovian atmosphere, as revealed for example by the images of Ortiz {\it et al.}\ (1998). However, it demonstrates the need to invoke spatial heterogeneity and the fact that most of the 5-\micron\ spectrum samples the vertical distributions of water and ammonia in the hot spots rather than on the whole planet. The radiance in our model of the hot spot region at 5\,\micron, $\sim$3.5 erg\,s$^{-1}$\,cm$^{-2}$\,sr$^{-1}$/\cm, is found comparable to (and not surprisingly higher than) that observed in natural hot spots, typically $\sim$1.5 erg\,s$^{-1}$\,cm$^{-2}$\,sr$^{-1}$/\cm (Carlson {\em et al.}\ 1993; Roos-Serote {\em et al.}\ 1998). In addition, the Nephelometer experiment on the board the Galileo probe indicated extinction optical depth less than 2.

Neglecting scattering is less important in this spatially heterogeneous approach than in the spatially homogeneous. Indeed, essentially no photons are emitted from below the cloud in the dark region. Introducing multiple scattering would then only change the cloud parameters required to make this region dark. The hot spot region was defined as cloud free, making the problem of scattering irrelevent there. Even if cloud optical depths have typical values of $\sim2$ in hot spots, the number of scattering events is low, and therefore the mean optical path only sligthly increased. Our retrieved ammonia abundances should then be reduced only by a few percents, leaving the ammonia vertical profile nearly unchanged.
%}

It is especially interesting to use our composite model to compare the ISO-SWS data with the Galileo probe measurements. The probe entered a hot spot, where it found a dry atmosphere. Two different experiments on board the probe have measured the ammonia vertical distribution. Folkner {\em et al.}\ (1998) analyzed the attenuation of the probe radio signal. They found the \nh\ mixing ratio to increase with depth from highly depleted at 0.4\,bar to $3.2\times 10^{-4}$ at the 4-bar level and $9\times 10^{-4}$ (${\rm 4\times solar}$) at 8\,bars, then roughly constant at deeper levels. Sromovsky {\em et al.}\ (1998), using the Net Flux Radiometer, found a similar profile down to 4\,bar, hence increasing with depth, but vertically constant at greater pressures at a mixing ratio of ${\rm 2.8\times 10^{-4}}$ (${\rm 1.3\times solar}$). These profiles are very similar to the one we derived from the ISO-SWS observations (Fig.~8). Therefore we tested the respective profiles of Sromovsky {\em et al.}\ (1998) and of Folkner {\em et al.}\ (1998) against the ISO-SWS data themselves (Fig.~10). To do so, since the mixing ratio derived by Folkner {\em et al.}\ (1998) at 1.5\,bar suffers from huge uncertainties, we modified this measurement within the given error bars, in order to get a good agreement with the observed spectrum at 5.3\,\micron.

The differences induced by the two profiles in the calculations are only marginal. The \nh\ spectral features (4.82, 4.92 and 4.94\,\micron) calculated for the Folkner {\em et al.}\ profile appear slightly deeper than in the observed spectrum. Yet, these lines lie in the neighborhood of phosphine or water absorptions, thus lowering the sensitivity to the \nh\ mixing ratio. Moreover, mismatches between the observed and the synthetic spectra, already mentioned previously, exist at some other wavelengths (4.86, 5.07 and 5.13\,\micron\ for example). %{\bf
The line at 5.07\,\micron\ in particular is purely due to phosphine, suggesting that our \ph\ mixing ratio is overestimated. However, reducing it degrades the fit in the center of the phosphine band at 4.8\,\micron. We leave this problem of the \ph\ mixing profile to future work, where constraints from the entire 4.4--4.8 \micron\ range will also be considered.
%}
The disagreements noted above are larger than the differences between the calculated spectra for the Sromovsky {\em et al.}\ and Folkner {\em et al.}\ profiles. Taking all of these sources of error leads to an estimation of the \nh\ mixing ratio of $(3.5^{+5.5}_{-1.3})\times10^{-4}$ at 7\,bars. An accurate determination of the deep ammonia abundance would therefore require a higher spectral resolution, in order to isolate the \nh\ lines, and a better accuracy of the spectroscopic data. In summary, we find that the ISO-SWS and the Galileo probe data are consistent with each other down to the 4-bar level, but that the error bar affecting the deep \nh\ mixing ratio from the ISO-SWS data is too large to discriminate between the two Galileo profiles.\\

\noindent {\em Absence of H$_2$O cloud}\\

Thermodynamical equilibrium models of Jupiter's atmosphere predict the presence of a water cloud around 5\,bars (Carlson {\em et al.}\ 1987). Such a cloud was invoked by Carlson {\em et al.}\ (1993) from Voyager/IRIS data. But its detection was not confirmed by the Galileo mission, either from Probe data (Sromovsky {\em et al.}\ 1998; Ragent {\em et al.}\ 1998), or from remote sensing observations with the Near-Infrared Mapping Spectrometer (Roos-Serote {\em et al.}\ 1998). Moreover, Roos-Serote {\em et al.}\ (1999) suggested that the contradiction between the Voyager/IRIS and Galileo/NIMS analyses might be a consequence of a calibration problem affecting the Voyager/IRIS data at wavelengths shorter than 5\,\micron.  We now show that a thick water cloud is also inconsistent with the ISO/SWS data. Here, however, the argument is based on the spectral region longward of 5.3\,\micron, a region which was not observed by Galileo/NIMS and which suffered from too low a signal-to-noise ratio in Voyager/IRIS data to be usable.
Figure~12 displays the synthetic spectrum computed with our composite model in the presence of a thick deep cloud. The region with the 1.0--1.9 bar cloud is unchanged (anyway it is not sensitive to the presence of a deep \ho\ cloud), while a cloud layer based at 5\,bars is introduced in the hot spot region. Its transmission is arbitrarily fixed to ${\rm T_c=0.5}$ and its scale height to ${\rm H_p=0.20\times H_g}$ (Carlson {\em et al.}\ 1993). All other parameters being fixed, both the continuum level and the \ho\ line depths are lowered by the presence of the \ho\ cloud. Thus the surface covering factor of the hot spot region (13\%)
is adjusted to fit the data at 5.05\,\micron, and the water mixing ratio is increased to 8\%
of saturation, in order to recover the observed contrast in the water lines. This model leads to a strong disagreement with the observed spectrum longward of 5.2\,\micron. The reason for this behavior is the following: the water cloud opacity essentially affects the center of the 5-\micron\ window, where half of the flux originates from levels deeper than 5\,bars. On the contrary, in the edges of the window, the thermal emission mainly comes from pressure levels lying between 2 and 5\,bars, thus unaffected by the water cloud. As a result the flux around 5.3\,\micron\ is enhanced with respect to the center the 5-\micron\ window and, for a 13\% surface covering factor of the hot spots, is inconsistent with the observations. This excess emission cannot be reduced in any way. The cloud emission, which now contributes only at a level of 10\% to the calculated flux, cannot be tuned to reduce the excess emission. We also tested the effect of the cutoff to the Lorentz profile (20\,\cm), applied to describe the sub-Lorentzian lineshapes. We found that it does not significantly affect the synthetic spectrum. Therefore we must reject the presence of an opaque cloud around 5\,bars, at least in the hot spots, in agreement with Galileo observations (Sromovsky {\em et al.}\ 1998; Ragent {\em et al.}\ 1998; Roos-Serote {\em et al.}\ 1998, 1999).

\subsection{${\rm\bf ^{15}N/^{14}N}$ isotopic ratio}

\indent\indent In the computation of the synthetic spectrum displayed on Fig.~2 we assumed a terrestrial value for the \ratio\ isotopic ratio (${\rm (^{15}N/^{14}N)_{\oplus}=3.68\times 10^{-3}}$). But, investigating the \inh\ absorptions at 9.60, 9.79, 9.97, 10.60, 11.07 and 11.31\,\micron, it appears clearly that the terrestrial isotopic ratio overestimates the abundance of \inh\ in Jupiter's atmosphere. As shown on Fig.~13, a \inh\ depletion of $\delta=-50\%$ is in good agreement with the ISO-SWS observations, where $$\rm\delta=100\times\lbrack{(^{15}N/^{14}N)_J\over(^{15}N/^{14}N)_{\oplus}}-1\rbrack.$$

This is at face value a very surprising result; however many parameters affect the retrieved \ratio\ value, leading to large uncertainties. The \inh\ lines probe the \inh\ mixing ratio at a given pressure ($\sim400$\,mbar). The deduced \ratio\ thus depends on the \rnh\ mixing ratio at the same pressure level. Therefore, in order to evaluate the range of the possible \ratio\ ratios, we must first take into account all the parameters which can modify the retrieved \rnh\ and \inh\ mixing ratios, and second simultaneously determine the \rnh\ and \inh\ mixing ratios at similar pressure levels. To meet this last point, the most suitable way is to simultaneously analyze a \rnh\ line and a \inh\ line of similar intensities, formed on the same continuum level. Therefore, in the following analysis, we will compare the weak line of the Q branch of the $\nu_2^s$ band of \rnh, located at 10.88\,\micron, with the \inh\ line at 11.07\,\micron.

The \rnh\ line has an intensity at 296\,K of $4.09\times 10^{-20}$\,cm\,molecule$^{-1}$ and an energy level of 721.134\,\cm. The line of \inh\ is in fact a triplet with respective intensities of $2.48\times 10^{-22}$, $3.90\times10^{-22}$ and $8.62\times10^{-22}$\,cm\,molecule$^{-1}$ at 296\,K for a terrestrial isotopic ratio, and energy levels of 104.894, 115.104 and 118.837\,\cm. At 132\,K, the temperature at 400\,mbar, the \inh\ triplet is thus intrinsically 1.4 times stronger than the \rnh\ line (the relative accuracy in intensity between lines of this strength is expected to be better than 10\%, including the uncertainties on the terrestrial \ratio). Yet, Fig.~14 shows that the line at 10.88\,\micron\ is observed to be deeper than the \inh\ triplet. Furthermore, the two lines lie on the same continuum level, which can be fitted simultaneously with the same \nh\ distribution and cloud transmission (Fig.~13). Therefore we think that the observed relative depths of the \inh\ and \rnh\ absorptions can only be explained by a significant depletion in \inh\ with respect to the terrestrial ratio.

We have tested the robustness of this conclusion and derived a \ratio\ range by varying all model parameters which can affect the retrieved \rnh\ and \inh\ mixing ratio profiles. These parameters are the following:
\begin{itemize}
\item The transmission of the cloud layer is subject to uncertainties which can change the inferred ammonia abundance: for a lower cloud transmission the continuum level is increased while the strongest \nh\ lines remain unaffected. Thus a lower ammonia mixing ratio is needed below 400\,mbar. We take this effect into account by allowing the cloud transmission to vary between 0.6 and 1.0 by step of 0.1. Transmissions lower than 0.6 are unrealistic and rejected, following the analyses of Gierasch {\em et al.}\ (1986) and Griffith {\em et al.}\ (1992).

\item The precise shape of the far wings of the \nh\ lines is not very well known, and a significant departure from the Lorentz profile is possible. The effect on the ammonia mixing ratio is that a sublorentzian profile requires, to reproduce the observed continuum, a larger \rnh\ content than a Lorentz profile. We take into account this effect by calculating the \ratio\ for four different line shapes. Two of them are Lorentz profiles with cutoffs at 20 and 50\,\cm\ respectively, while the other two, following results obtained by Burch {\em et al.}\ (1969) on the CO$_2$ molecule, are Lorentz profiles multiplied respectively by shape factors $\chi(\nu)=e^{-(\nu-\nu_0)/80}$ and $\chi(\nu)=e^{-(\nu-\nu_0)/30}$, where $\nu-\nu_0$ is the offset from the line center in \cm.

\item The temperature profile is the parameter which produces the most important variations in the ammonia mixing ratio. Following our analysis of Section 3.1.1, we allowed it to vary among three possible profiles: the nominal one, a warmer one ($+2$\,K) and a colder one ($-2$\,K).
\end{itemize}

For each combination of the three temperature profiles, of the four line shape factors and of the five different cloud transmissions, we retrieved the ammonia vertical profile which fits the 10.88-\micron\ line and the surrounding continuum (i.e.\ 60 different profiles). We then determined the \inh\ mixing ratio which best fits the 11.07-\micron\ line. The results are summarized in Table~II. The influence of the cloud transmission and of the ammonia line shape factor on the \ratio\ is relatively weak. The reason is that variations of these two parameters mainly lead to variations of the ammonia mixing ratio at 630\,mbar but only slightly at 400\,mbar, the level probed by the \inh\ lines. The change in the temperature profile alters the retrieved \ratio\ ratio more significantly.

However, none of the inferred values for the \ratio\ ratio is in agreement with the terrestrial value. A terrestrial value could be obtained only for a temperature profile 8\,K colder than the nominal profile, leading to a temperature as low as 100\,K at the tropopause. Such a situation is very unlikely and, moreover, leads to a totally unacceptable fit of the whole 10-\micron\ region (Fig.~15). We therefore conclude from the analysis of the 10.07-\micron\ line that the \ratio\ at 400\,mbar is subterrestrial with a value of $\delta=(-50^{+25}_{-25})\%$. This result is qualitatively confirmed by the other \inh\ lines (Fig.~13).

\section{Discussion}

\subsection{Ammonia vertical profile}

\indent\indent We find an ammonia mixing ratio at the 300\,mbar level of $3.2\times10^{-7}$. This value is lower than the one inferred from the Voyager/IRIS observations by Kunde {\em et al.}\ (1982) ($1.4\times10^{-6}$ in the NEB), and by Griffith {\em et al.}\ (1992) ($1.4\times10^{-6}$ in the STrZ). On the contrary, Carlson {\em et al.}\ (1993), also from Voyager/IRIS data, found in the NEB an ammonia mixing ratio ($2.4\times10^{-7}$) similar to the one we deduced. These differences may be partly related to spatial variations, but are more likely the consequence of differences in the model parameters, such as the location of the distribution anchor points, or in the temperature profiles. Lara {\em et al.}\ (1998), from high spectral resolution observations, investigated spatial variations of the ammonia mixing ratio at 240\,mbar. They found it to range between $4.8\times10^{-8}$ and $1.0\times10^{-9}$, the largest values being clustered at $10^{\circ}$S--$18^{\circ}$S\,latitude,
and the lowest at $30^{\circ}$S--$35^{\circ}$S\,latitude. Their interpretation of this difference is that the eddy diffusion coefficient (K) is higher at $10^{\circ}$S--$18^{\circ}$S (${\rm K\approx4000\,cm^2\,s^{-1}}$) than at $30^{\circ}$S--$35^{\circ}$S, where they inferred a value of ${\rm K\le400\,cm^2\,s^{-1}}$. Even if our analysis does not probe the same pressure level, we can compare our results with those of Lara {\em et al.}\ (1998), since the calculated eddy coefficient is essentially not determined by the \nh\ mixing ratio at a specific level but rather by the \nh\ mixing ratio scale height (${\rm H_q}$). Inspection of their Fig.~7 shows that at $10^{\circ}$S--$18^{\circ}$S ${\rm H_q=0.1\times H_g}$, while ${\rm H_q=0.05\times H_g}$ at $30^{\circ}$S--$35^{\circ}$S. Our inferred value is ${\rm H_q=0.085\times H_g}$. The different anchor points assumed in both studies can greatly affect the retrieved scale height, but this suggests that the high value of the eddy diffusion coefficient (i.e.\ ${\rm K\approx4000\,cm^2\,s^{-1}}$) represents the average situation over Jupiter's low latitude. As already noted by Lara {\em et al.}\ (1998) this finding must be linked with the global pattern of the ortho-para hydrogen ratio (Conrath and Gierasch 1984), which exhibits a departure from the equilibrium maximum at the equator and decreasing toward the poles.
This departure can also result from a stronger eddy coefficient at low latitudes than at the poles (Conrath and Gierasch 1984). However, this result is in contradiction with the work of Edgington {\em et al.}\ (1998). They combined a photochemical model with HST/FOS observations of Jupiter to retrieve the \nh\ mixing ratio and the eddy diffusion coefficient at several latitudes between 6\degre\,N and 48\degre\,N. They found no latitudinal variations at pressures greater than 170\,mbar. Reasons for this disagreement with Lara {\em et al.}\ (1998) are unclear, since the Rayleigh scattering allows the photons to penetrate as deep as the 300-mbar level at 220\,nm, and forbid a consistent picture of the photochemical and dynamical processes taking place in the upper troposphere of Jupiter.

At pressures greater than 300\,mbar, we find a mixing ratio of $1.8\times10^{-5}$ at 450\,mbar and $4.2\times10^{-5}$ at 600\,mbar. This is in satisfactory agreement with previous measurements. Carlson {\em et al.}\ (1993) and Kunde {\em et al.}\ (1982) found respectively $8.8\times10^{-6}$ and $8.9\times10^{-6}$ at 450\,mbar, and $2.8\times10^{-5}$ and $3.0\times10^{-5}$ at 600\,mbar, while Griffith {\em et al.}\ (1992) inferred in the STrZ $1.4\times10^{-5}$ at 450\,mbar and $3.5\times10^{-5}$ at 600\,mbar respectively. The values of Carlson {\em et al.}\ (1993) and Kunde {\em et al.}\ (1982) are lower than the value of Griffith {\em et al.}\ (1992) and ours, probably because they specifically probe the NEB, a region of low relative humidity.

In the 5-\micron\ window, we find that a cloud layer is needed to attenuate the flux in agreement with many other earlier studies. This cloud layer must have an effective brightness temperature lower than 166\,K. This contradicts the previous analyses from Voyager and KAO spectra (B\'ezard {\em et al.}\ 1983; Bjoraker {\em et al.}\ 1986; Lellouch {\em et al.}\ 1989) which were limited to the 4.8--5.2\,\micron\ range and assumed a cloud effective brightness temperature of $\sim$190--200\,K. A temperature of 166\,K can be obtained with various cloud models, from a deep, vertically extended layer to a thin layer located around the 1-bar level. It implies that the 5-\micron\ attenuating cloud could also contribute significantly to the opacity above the 1-bar level. The same conclusion was reached by Drossart {\em et al.}\ (1998) who found, from the coldest Galileo/NIMS spectra on the night side of the planet, a cloud effective brightness temperature of 160\,K or lower. In addition, we note that the 5-\micron\ attenuating cloud cannot be unambiguously identified with the expected \nhsh\ cloud layer. 

From the 5-\micron\ observations we find that the \nh\ mixing ratio increases downward at least down to 4\,bars and that a constant mixing ratio below the condensation level can be rejected. %{\bf
Although our simple model does not include scattering, we qualitatively find that multiple scattering can only increase this trend.
%}
This ammonia profile is associated with a highly subsaturated water profile. This very dry meteorological situation does not reflect the global state of the planet, but rather pertains to small areas, most probably the 5-\micron\ hot spots. This is strongly supported by the consistency between our inferred ammonia profile and the profiles of Sromovsky {\em et al.}\ (1998) and Folkner {\em et al.}\ (1998) determined from the Galileo probe measurements.

Previous analyses of the 5-\micron\ window have not put in evidence such a vertically variable distribution, and this for various reasons. Bjoraker {\em et al.}\ (1986) and Lellouch {\em et al.}\ (1989) did not focus on the vertical distribution of ammonia. Both of them assumed the distribution to follow the profile derived by Kunde {\em et al.}\ (1982) above 0.7\,bar, and to be constant below at a mixing ratio determined from a best fit to their observations. The spatial variations of the deep \nh\ abundance found by Lellouch {\em et al.}\ (1989) thus mostly reflect the variations of the average ammonia mixing ratio near the 2-bar level. On the other hand, from a reanalysis of the Voyager/IRIS data, Carlson {\em et al.}\ (1993) found an increasing ammonia abundance down to 2\,bars and constant at deeper levels in the NEB. We think that this is not in contradiction with our results since the IRIS resolution (4.3\,\cm), approximately 3 times lower than that of SWS ($\sim1.3$\,\cm), did not allow to resolve the \nh\ lines shortward of 5.3 \micron\, which probe the 4-bar pressure level.

From radio wavelength observations de Pater (1986) found that the ammonia mixing ratio must be constant at pressures greater than 2\,bar. Reason for this disagreement with the ISO-SWS observations and the Galileo probe measurements is not clear. Possible explanations are uncertainties on the subtraction of the Jupiter synchrotron radiation or on the ammonia lineshape factors (Moreno 1998).

The global N/H ratio of Jupiter, inferred from the \nh\ mixing ratio at 7\,bar suffers, as stated in Section 3.1.2, from very large uncertainties ($\rm[NH_3]/[H_2]=(3.5^{+5.5}_{-1.3})\times10^{-4}$, $\rm N/H=(1.6^{+2.4}_{-0.6})\times solar$). This leads to a C/N ratio of $\rm(1.8\pm1.1)\times solar$. Since the range encompasses the solar value, this does not allow us to confirm or refute the cometary origin of the planetesimals which have enriched Jupiter in heavy elements. However, the agreement between the ammonia profile inferred from the attenuation of the Galileo probe radio signal (Folkner {\em et al.}\ 1998) and the one retrieved from the ISO-SWS data, favors the high N/H (hence low C/N ($\sim$0.7), favoring a non-cometary origin) measured by Folkner {\em et al.}\ (1998), in contradiction with the centimetric observations of de Pater (1986) and Joiner and Steffes (1991).

\subsection{${\rm\bf ^{15}N/^{14}N}$ isotopic ratio}

\indent\indent The determination of the protosolar $^{15}$N/$^{14}$N isotopic ratio is still a debated question, since the Sun and the meteorites exhibit large variations in their isotopic composition of nitrogen. The solar value, measured in the solar wind trapped in the lunar regolith, shows secular variations from -21\% to +12\% (relative to the terrestrial value of ${\rm^{15}N/^{14}N=3.68\times10^{-3}}$) (Geiss and Bochsler, 1982). The \ratio\ value in meteorites (reviewed by Pillinger, 1984), varies from -32\% to +21\%, but the majority of the samples clusters between -3\% and +4\%. Different scenarios have been proposed to account for these different values. The interpretation of Thiemens and Clayton (1981) yields a protosolar value of -3\%, the lighter nitrogen component in meteorites being due to the admixture of an alien $^{14}$N-rich phase, and the heavier nitrogen to fractionation. Kerridge (1980) proposed that the protosolar value is $<$-21\% and that the $^{15}$N-enrichment in meteorites and planets is achieved by fractionation. Finally Geiss and Bochsler (1982) proposed a protosolar value of +12\%, and that the meteoritic ratios are the result of the admixture of a light nitrogen alien phase.

No definitive argument has yet been advanced to discriminate between the three scenarios, and important issues for the formation of the Solar System, such as the degree and temperature of fractionation in meteorites, or the degree of mixing with interstellar grains, are left unanswered. Therefore Kerridge (1982) had hoped that the Galileo probe could resolve this question. He argued that the greater part of nitrogen in Jupiter should not have suffered isotopic fractionation and that its $^{15}$N/$^{14}$N ratio should be regarded as protosolar. Unfortunately, isotopic ammonia has blent with other chemical species in the Galileo probe Mass Spectrometer, thus precluded any reliable measurement of the \ratio\ ratio. Therefore our hopes still rely on remote sensing observations of Jupiter.

Previous determinations of the \ratio\ isotopic ratio have been made by Encrenaz {\em et al.}\ (1978) and Tokunaga {\em et al.}\ (1980), who found a \ratio\ ratio consistent with the terrestrial value, but within a factor of 2. These results thus largely differ from our measurement of $(\delta=-50^{+25}_{-25})$\,\%. As we have shown in Section 3.2, an accurate determination of the \ratio\ ratio requires an accurate modeling of both \inh\ and \rnh\ at the same pressure level. This can only be done using the \rnh\ line at 10.88\,\micron\ and the \inh\ line at 11.07\,\micron. Tokunaga {\em et al.}\ (1980) measured the full spectrum of Jupiter from 10 to 12\,\micron, allowing the comparison of the \inh\ and \rnh\ lines. Investigation of their Fig.~2, where the signal to noise ratio is the highest, shows that the brightness temperatures at 10.88 and 11.07\,\micron\ are almost identical (135\,K). This agrees with the ISO-SWS data, but it is in contradiction with a terrestrial \ratio\ ratio (Section 3.2). We are thus inclined to conclude that the spectrum of Tokunaga {\em et al.}\ (1980) also favors a depletion in \inh. Even though they focussed their analysis on \rnh\ and \ph, Lara {\em et al.}\ (1998) also observed the \inh\ line at 943\,\cm\ (10.60\,\micron). Inspection of their Fig.~8 shows that the terrestrial value of \ratio\ overestimates the depth of the \inh\ line, the \rnh\ mixing ratio being determined at 380\,mbar from the \rnh\ absorption at 921\,\cm\ (10.86\,\micron). The two lines do not exactly lie on the same continuum level, but this nevertheless favors again a depletion in \inh.

Hence, even if a systematic problem in the analysis of the $\nu_2$ band of ammonia cannot be totally ruled out, several spectroscopic observations support a depletion of Jupiter in $\rm{^{15}N}$. A first hypothesis is that the observed \ratio\ depletion reflects the global isotopic ratio (T.\ Owen, priv.\ comm.; paper in preparation). This would imply that nitrogen in Jupiter is not of cometary origin, since comets apparently have a terrestrial \ratio\ (as measured in Hale-Bopp by Jewitt {\em et al.}\ (1997)). This conclusion would agree with the Galileo finding that Jupiter's C/N is not oversolar. In the case of an origin of nitrogen from the gaseous protosolar nebula, both the Sun and, presumably, the interstellar medium should exhibit a similar \ratio\ as Jupiter (T.\ Owen, priv.\ comm.). We note, however, that in this case the jovian N/H should be solar, which does not seem to be the case. Reconciling the \ratio\ and N/H measurements in Jupiter would in fact imply an origin from planetesimals enriched in $^{14}$N and depleted in \ratio\ compared to the solar value. The nature and fate of these bodies however is essentially speculative at this point. 

These difficulties to reconcile the observed \ratio\ with the abundance in other solar system objects raise the question whether or not the observed \ratio\ ratio is representative of the bulk composition of the planet. Indeed we measured the \inh\ mixing ratio at a pressure level ($\sim400$\,mbar) where ammonia experiences condensation in \nh-ice particles and photodissociation from the solar UV radiation, two processes which can plausibly alter the \ratio\ ratio. 

The photodissociation of ammonia is dominated by the \~A--\~X band system at 200\,nm (Atreya 1986), the unit optical depth being reached at $\sim$200\,mbar in the strongest bands. Since \ratio\ ratio is at most $5\times10^{-3}$, the \inh\ band system could remain optically thin at pressures higher than that of \rnh, the photodissociation of \inh\ hence taking place below the \rnh\ photodissociation level. This process could alter the \ratio\ at 400\,mbar. However, it requires the \rnh\ \~A--\~X system to be transparent to \inh\ absorptions. In turn it requires either the \inh\ and \rnh\ systems to be shifted with respect to each other by a large offset, or the individual rotational lines constituting the bands to be spaced by steps larger than their intrinsic widths. None of these conditions appear to be met (Douglas 1963). The shift between \inh\ and \rnh\ bands is only $\sim$30\,\cm\ ($\sim$1\,\AA), less than the half-width of each individual band, and the \rnh\ bands are too diffuse to let photons pass trough their rotational structure. The \inh\ photodissociation thus takes place at the same pressure level as that of \rnh, and is unlikely to produce any significant alteration of the \ratio\ ratio at 400\,mbar.

Perhaps more promising is the possibility of isotopic enrichment of ${\rm^{14}N}$ due to condensation. Indeed, it has long been recognized that the vapor pressure of a molecule depends on its isotopic composition, generally leading to an enrichment of the condensed phase in the heavy isotope. Such an isotopic fractionation is widely documented for the condensation of water in the terrestrial atmosphere, where it is known to be an important process. At 0\degre\,C the fractionation coefficient $\alpha$, which measures the enrichment of the liquid phase with respect to the vapour phase at thermodynamical equilibrium, is 1.123 for D and 1.0177 for $^{18}$O (Majoube, 1971a). The deuterium gradient during the condensation of water can thus be of the order of a few tens of \%\,C$^{\circ-1}$ (Jouzel, 1986) and D/H ratios as low as $0.4\times{\rm (D/H)_{SMOW}}$ have been measured in the stratosphere (Rinsland {\em et al.}\ 1984) (SMOW stands for Standard Mean Ocean Water).

A similar process can take place on Jupiter during the condensation of gaseous ammonia in \nh-ice particles. We have calculated the fractionation coefficient $\alpha$ needed to reproduce our observed \ratio. We used an ``open cloud'' model, where the condensed phase is assumed to be immediately removed after its formation and to leave the atmospheric parcel in isotopic equilibrium with the vapour phase. This results in a continuous decrease of the heavy isotope content, which is simply written, following Jouzel (1986) by: $${{\rm d}\delta_V\over1+\delta_V}=(\alpha-1){{\rm d}n_V\over n_V},$$ where $n_V$ is the mixing ratio of ammonia and $\delta_V$ the enrichment or depletion in \inh\ of the vapour phase with respect to the terrestrial value. If we assume a constant fractionation coefficient--- i.e.\ independent of the temperature --- we can integrate the preceding equation to: $$\delta_V=(1+\delta_0)\left({n_V\over n_0}\right)^{\alpha-1}-1,$$ where $\delta_0$ and $n_0$ are respectively the isotopic content of the vapour phase and the ammonia mixing ratio at the condensation level. We assume $\delta_0=0$ (i.e.\ that the \ratio\ ratio is terrestrial below the cloud), and we solve for $\alpha$ to match the observed $\delta_V$.

The results for various atmospheric models are displayed in Table~II. The explanation of the observed \ratio\ isotopic ratio by an isotopic fractionation during the cloud condensation requires high fractionation coefficients ([1.08,1.45]). This coefficient has not been measured, as far as we know, for the solid-vapour transition of \nh, but experiments have been carried for the liquid-vapour transition (Thode, 1940). At 200\,K $\alpha$ equals 1.005, well below our lowest value of $\alpha=1.08$. Comparisons can also be made with the vapour/solid equilibrium for H$_2$$^{18}$O, since the $^{15}$N and $^{18}$O atoms play similar central roles in their respective molecules. The fractionation coefficient of H$_2$$^{18}$O is also low, $\alpha=1.0152$ at 0\degre\,C (Majoube, 1971b). In addition, our ``open cloud'' model most likely underestimates the needed fractionation coefficient since the heavy isotope is constantly removed from the cloud through precipitation.

However, the vapour pressure isotope effect can present a strong temperature dependence. Therefore in the absence of laboratory data for the vapour/solid transition of ammonia at the relevant temperatures, this effect remains a possible explanation of the low observed \ratio\ ratio. In addition, the needed fractionation coefficient would be lower if the expected \nhsh\ cloud condensation had already reduced the \ratio\ ratio priot to \nh\ condensation in an upwelling column of gas.

\section{Conclusions}

\indent\indent We have analyzed the ISO-SWS spectrum  of Jupiter in terms of the ammonia vertical distribution. The observations at 5 and 10 \,\micron\ cannot be interpreted with a spatially homogeneous atmosphere. At 5\,\micron, the hot spots, covering 6.5\% of the SWS aperture, dominate the overall flux, the rest of the planet being covered by an optically thick cloud layer. At 10\,\micron, the sounded atmospheric levels are limited to pressures less than 1\,bar. The thermal contrast is thus much lower than at 5\,\micron\ %{\bf
since the overall flux is dominated by the portion that emerges from the large fraction of the planet that is not within 5-\micron\ hot spots.
%}
In this region we find a \nh\ vertical distribution close to solar around 1\,bar, then following a subsaturated profile before undergoing depletion by photodissociation above 400\,mbar. The hot spots are found very depleted in condensible species. The \nh\ mixing ratio increases with depth at least down to the 4-bar level, well below the condensation level. The ammonia distribution is associated with a low water relative humidity ($<2\%$). These very dry conditions are similar to those inferred both from the in-situ measurements of the Galileo probe (Folkner {\em et al.}\ 1998; Niemann {\em et al.}\ 1998) and from NIMS data analysis (Roos-Serote {\em et al.}\ 1998).

The measured ammonia mixing ratio at 7\,bars ($\rm[NH_3]/[H_2]=(3.5^{+5.5}_{-1.3})\times10^{-4}$, $\rm N/H=(1.6^{+2.4}_{-0.6})\times$ solar) suffers from large uncertainties. Yet, the qualitative agreement of the ISO-SWS and Galileo probe ammonia vertical distributions at pressures lower than 4 bars strengthens the high N/H value (${\rm(4\pm0.5)\times solar}$) found by Folkner {\em et al.}\ (1998). Such a value, implying that C/N is not supersolar, points to an origin else than cometary for the planetesimals which have enriched Jupiter in heavy elements.

We find that a thick water cloud around the 5-bar level is not compatible with the ISO-SWS data longward of 5.3\,\micron. The same conclusion was reached independently by Ragent {\em et al.}\ (1998) using the Nephelometer on board the Galileo probe and by Roos-Serote {\em et al.}\ (1998, 1999) who analyzed the slopes of the NIMS spectra between 4.7 and 5.2\,\micron. 

We find, from the analysis of the $\nu_2$ band of \inh\, a \ratio\ ratio at 400\,mbar essentially half of the terrestrial ratio (${\rm (^{15}N/^{14}N)_{J}=(1.9^{+0.9}_{-1.0})\times10^{-3}}$, while\\
 ${\rm (^{15}N/^{14}N)_{\oplus}=3.68\times 10^{-3}}$). So far, such a value is unique in the solar system. If it reflects the global jovian \ratio, it would imply that Jupiter's nitrogen does not originate from comets. However, an origin from the solar nebula gas may be inconsistent with the N/H observed by the Galileo probe. Alternatively, it could be the consequence of an isotopic fractionation which takes place above the ammonia cloud base. A possible process is the isotopic fractionation occuring during ammonia condensation. We have calculated that the jovian \ratio\ could be explained if the condensed phase is enriched in ${\rm^{15}N}$ at thermodynamical equilibrium by at least a factor of 1.08. This value seems high compared to the other known isotope fractionation coefficients. Nevertheless, in the absence of laboratory data specific to \nh, we cannot derive any firm conclusions. If this process really occurs on Jupiter, it could be a very important tracer of the cloud microphysics. Laboratory measurements of the fractionation coefficient at the temperatures relevant to the jovian atmosphere would thus be very useful.

\newpage
\section*{Acknowledgements}

\indent\indent This study is based on observations with ISO, an ESA project with instruments funded by ESA Member States (especially the principal investigators countries: France, Germany, the Netherlands and the United Kingdom), and with participation of ISAS and NASA. We thank C.\ A.\ Griffith, A. Armengaud and D. Gautier for useful discussions.

\newpage
\section*{References}

\noindent Anders, E., and N.\ Grevesse 1989. Abundances of the elements:
meteoritic and solar. {\it Geochim.\ Cosmochim.\ Acta} {\bf 53}, 197--214.\\

\noindent Atreya, S.\ K.\ 1986. {\it Atmospheres and ionospheres of the outer
planets and their satellites}. Springer-Verlag, Berlin.\\

\noindent Banfield, D., P.\ J.\ Gierasch, M.\ Bell, E.\ Ustinov,
A.\ P.\ Ingersoll, A.\ R.\ Vasavada, R.\ A.\ West, and M.\ J.\ S.\ Belton
1998. Jupiter's Cloud Structure from Galileo Imaging Data. {\it Icarus}
{\bf 135}, 230--250.\\

\noindent B\'ezard, B., J.-P.\ Baluteau, and A.\ Marten 1983. Study of the deep
cloud structure in the equatorial region of Jupiter from Voyager infrared and
visible data. {\it Icarus} {\bf 54}, 434--455.\\

\noindent Birnbaum, G., A.\ Borysow, and G.\ S.\ Orton 1996. Collision-induced
absorption of H$_2$-H$_2$ and H$_2$-He in the rotational and fundamentals bands
for planetary applications. {\it Icarus} {\bf 123}, 4--22.\\

\noindent Bjoraker, G.\ L., H.\ P.\ Larson, and V.\ G.\ Kunde 1986. The gas
composition of Jupiter derived from 5-\micron\ airborne spectroscopic
observations. {\it Icarus} {\bf 66}, 579--609.\\

\noindent Borysow, J., L.\ Trafton, L.\ Frommhold, and G.\ Birnbaum 1985.
Modeling of pressure induced far infrared absorption spectra: molecular
hydrogen pairs. {\it Astrophys.\ J.}\ {\bf 296}, 644--654.\\

\noindent Borysow, J., L.\ Frommhold, and G.\ Birnbaum 1988. Collision induced
rototranslational absorption spectra of ${\rm H_2}$-He pairs at temperatures
from 40 to 3000\,K. {\it Astrophys.\ J.}\ {\bf 326}, 509--515.\\

\noindent Brooke, T.\ Y., R.\ F.\ Knacke, T.\ Encrenaz, P.\ Drossart,
D.\ Crisp, and H.\ Feuchtgruber 1998. Models of the ISO 3-$\mu$m reflection
spectrum of Jupiter. {\it Icarus} {\bf 136}, 1--13.\\

\noindent Brown, L.\ R., and D.\ B.\ Peterson 1994. An empirical expression for
linewidths of ammonia from far-infrared measurements. {\it J.\ Mol.\
Spectrosc.}\ {\bf 168}, 593--606.\\

\noindent Brown, L.\ R., and C.\ Plymate 1996. H$_2$-broadened H$_2$$^{16}$O in
4 infrared bands between 55 and 4045 \cm. {\it J.\ Quant.\ Spectrosc.\ Radiat.\
Trans.}\ {\bf 56}, 263--282.\\

\noindent Burch, D.\ E., D.\ A.\ Gryvnak, R.\ R.\ Patty, and C.\ E.\ Bartky
1969. Absorption of infrared radiant energy by CO$_2$ and H$_2$. IV.\ Shapes of
collision-broadened CO$_2$ lines. {\it J.\ Optic.\ Soc.\ Americ.}\ {\bf 59},
267--280.\\

\noindent Carlson, B.\ E., M.\ J.\ Prather, and W.\ B.\ Rossow 1987. Cloud
chemistry on Jupiter. {\it Astrophys.\ J.}\ {\bf 322}, 559--572.\\

\noindent Carlson, B.\ E., A.\ A.\ Lacis, and W.\ B.\ Rossow 1993. Tropospheric
gas composition and cloud structure of the jovian North Equatorial Belt.
{\it J.\ Geophys.\ Res.}\ {\bf 98}, 5251--5290.\\

\noindent Carlson, B.\ E., A.\ A.\ Lacis, and W.\ B.\ Rossow 1994. Belt-zone
variations in the jovian cloud structure. {\it J.\ Geophys.\ Res.}\ {\bf 99},
14623--14658.\\

\noindent Conrath, J.\ B., and P.\ J.\ Gierasch 1984. Global variation of the
para hydrogen fraction in Jupiter's atmosphere and implications for dynamics of
the outer planets. {\it Icarus} {\bf 57}, 184--204.\\

\noindent Conrath, J.\ B., and P.\ J.\ Gierasch 1986. Retrieval of ammonia
abundances and cloud opacities on Jupiter from Voyager IRIS spectra.
{\it Icarus} {\bf 67}, 444--455.\\

\noindent Dana, V., J.-Y.\ Mandin, G.\ Tarrago, W.\ B.\ Olson, and B.\ B\'ezard
1993. Absolute infrared intensities in the fundamentals $\nu_1$ and $\nu_3$ of
arsine. {\it J.\ Molec.\ Spectrosc.}\ {\bf 159}, 468--480.\\

\noindent de Graauw, T., and 61 colleagues 1996. Observing with the ISO
Short-Wavelength Spectrometer. {\it Astron.\ Astrophys.}\ {\bf 315},
L49--L54.\\

\noindent de Pater, I.\ 1986. Jupiter's zone-belt structure at radio
wavelengths. II.\ Comparison of observations with model atmosphere
calculations. {\it Icarus} {\bf 68}, 344--365.\\

\noindent Douglas, A.\ E.\ 1963. Electronically excited states of ammonia.
{\it Discuss.\ Faraday Soc.}\ {\bf 35}, 158--174.\\

\noindent Drossart, P., J.-P.\ Maillard, M.\ Roos-Serote, E.\ Lellouch,
and T.\ Encrenaz 1996. High resolution spectroscopy of the GRS and NEB hot
spots of Jupiter following Galileo observations. {\it Bull.\ Amer.\ Astron.\
Soc.}, {\bf 28}, 223.\\

\noindent Drossart, P., and 11 colleagues 1998. The solar reflected component
in Jupiter's 5-$\mu$m spectra from NIMS/Galileo observations. {\it J.\
Geophys.\ Res.}, {\bf 103}, 23043--23049.\\

\noindent Eberhardt, P.\ 1998. Composition of comets: the in situ view. In
{\it Cometary nuclei in space and time} (M.\ A'Hearn, Ed.). Atron.\ Soc.\ of
the Pacific Conf.\ Series.\\

\noindent Edgington, S.\ G., S.\ K.\ Atreya, L.\ M.\ Trafton, J.\ J.\ Caldwell,
R.\ F.\ Beebe, A.\ A.\ Simon, R.\ A.\ West, and C.\ Barnet 1998. On the
latitude variation of ammonia, acetylene, and phosphine altitude profiles on
Jupiter from HST Faint Object Spectrograph observations. {\it Icarus}
{\bf 133}, 192--209.\\

\noindent Encrenaz, T., M.\ Combes, and Y.\ Z\'eau 1978. The spectrum of
Jupiter between 10 and 13 $\mu$: an estimate of the jovian ${\rm^{15}N/^{14}N}$
ratio. {\it Astron.\ Astrophys.}\ {\bf 70}, 29--36.\\

\noindent Encrenaz, T., and 19 colleagues 1996. First results of ISO-SWS
observations of Jupiter. {\it Astron.\ Astrophys.}\ {\bf 315}, L397--L400.\\

\noindent Folkner, W.\ M., R.\ Woo, and S.\ Nandi 1998. Ammonia abundance in
Jupiter's atmosphere derived from the attenuation of the Galileo probe's radio
signal. {\it J.\ Geophys.\ Res.}\ {\bf 103}, 22847--22855.\\

\noindent Gautier, D., B.\ J.\ Conrath, M.\ Flasar, R.\ Hanel, V.\ Kunde,
A.\ Chedin, and N.\ Scott 1982. The helium abundance of Jupiter from Voyager.
{\it J.\ Geophys.\ Res.}\ {\bf 86}, 8713--8720.\\

\noindent Geiss, J., and P.\ Bochsler 1982. Nitrogen isotopes in the solar
system. {\it Geochim.\ Cosmochim.\ Acta} {\bf 46}, 529--548.\\

\noindent Gierasch, P.\ J., B.\ J.\ Conrath, and J.\ A.\ Magalh\~{a}es 1986.
Zonal mean properties of Jupiter's upper troposphere from Voyager infrared
observations. {\it Icarus} {\bf 67}, 456--483.\\

\noindent Griffith, C.\ A., B.\ B\'ezard, T.\ Owen, and D.\ Gautier 1992. The
tropospheric abundances of NH$_3$ and PH$_3$ in Jupiter's Great Red Spot, from
Voyager IRIS observations. {\it Icarus} {\bf 98}, 82--93.\\

\noindent Hanel, R., and 14 colleagues 1979. Infrared observations of the
jovian system from Voyager 2. {\it Science} {\bf 206}, 952--956.\\

\noindent Husson, N., B.\ Bonnet, N.\ A.\ Scott, and A.\ Ch\'edin 1992.
Management and study of spectroscopic information: the GEISA program. {\it J.\
Quant.\ Spectrosc.\ Radiat.\ Transfer} {\bf 48}, 509--518.\\

%\noindent Izawa, Y., H.\ Noguchi, C.\ Yamanaka 1975. Isotope separation using
%the laser. Separation of nitrogen-14 and -15 by a two-step photodissociation
%of the ammonia molecule. {\it Nippon Genshiryoku Gakkaishi} {\bf 17}, 166--168
%(in Japanese).\\

\noindent Jewitt, D.\ C., H.\ E.\ Matthews, T.\ Owen, and R.\ Meier 1997.
Measurement of $^{12}$C/$^{13}$C, $^{14}$N/$^{15}$N, $^{32}$S/$^{34}$S ratios
in comet Hale-Bopp (C/1995 O1). {\it Science} {\bf 278}, 90--93.\\

\noindent Joiner, J., and P.\ G.\ Steffes 1991. Modeling of Jupiter's
millimeter wave emission utilizing laboratory measurements of ammonia (\nh)
opacity. {\it J.\ Geophys.\ Res.}\ {\bf 96}, 17463--17470.\\

\noindent Jouzel, J.\ 1986. Isotopes in cloud physics: multiphase and
multistage condensation process. In {\it Handbook of environmental isotope
geochemistry, the terrestrial environment, Vol.~2}, (P.\ Fritz, and
J.\ C.\ Fontes, Eds.), pp.\ 61--112. Elsevier, New-York.\\

\noindent Kerridge, J.\ F.\ 1980. Accretion of nitrogen during the growth of
planets. {\it Nature} {\bf 283}, 183--184.\\

\noindent Kerridge, J.\ F.\ 1982. Whence so much $^{15}$N? {\it Nature}
{\bf 295}, 643--644.\\

\noindent Kessler, M.\ F., and 10 colleagues 1996. The Infrared Space
Observatory (ISO) mission. {\it Astron.\ Astrophys.}\ {\bf 315}, L27--L31.\\

\noindent Kunde, V., R.\ Hanel, W.\ Maguire, D.\ Gautier, J.-P.\ Baluteau,
A.\ Marten, A.\ Ch\'edin, N.\ Husson, and N.\ Scott 1982. The tropospheric gas
composition of Jupiter's North Equatorial Belt (NH$_3$, PH$_3$, GeH$_4$,
H$_2$O) and the jovian D/H isotopic ratio. {\it Astrophys.\ J.}\ {\bf 263},
443--467.\\

\noindent Lara, L.-M., B.\ B\'ezard, C.\ A.\ Griffith, J.\ H.\ Lacy, and
T.\ Owen 1998. High-resolution 10-micronmeter spectroscopy of ammonia and
phosphine lines on Jupiter. {\it Icarus} {\bf 131}, 317--333.\\

\noindent Lellouch, E., N.\ Lacome, G.\ Guelachvili, G.\ Tarrago, and
T.\ Encrenaz 1987. Ammonia: experimental absolute linestrengths and
self-broadening parameters in the 1800- to 2100-\cm\ range.
{\it J.\ Mol.\ Spectrosc.}\ {\bf 124}, 333--347.\\

\noindent Lellouch, E., P.\ Drossart, and T.\ Encrenaz 1989. A new analysis of
the jovian 5-$\mu$m Voyager/IRIS spectra. {\it Icarus} {\bf 77}, 457--465.\\

\noindent L\'evy, A., N.\ Lacome, and G.\ Tarrago 1993. Hydrogen- and
helium-broadening of phosphine lines. {\it J.\ Mol.\ Spectrosc.}\ {\bf 157},
172--181.\\

\noindent Majoube, M.\ 1971a. Fractionnement en oxyg\`ene 18 et en deut\'erium
entre l'eau et sa phase vapeur. {\it J.\ Chim.\ Phys.\ Physicochim.\ Biol.}\
{\bf 68}, 1423--1436.\\

\noindent Majoube, M.\ 1971b. Fractionnement en oxyg\`ene 18 entre la glace et
la vapeur d'eau. {\it J.\ Chem.\ Phys.\ Physicochim.\ Biol.}\ {\bf 68},
625--636.\\

\noindent Margolis, J.\ S.\ 1996. H$_2$ and He broadened methane. {\it J.\
Quant.\ Spectrosc.\ Radiat.\ Trans.}\ {\bf 55}, 823--836.\\

\noindent Marten, A., D.\ Rouan, J.-P.\ Baluteau, D.\ Gautier, B.\ J.\ Conrath,
R.\ A.\ Hanel, V.\ Kunde, R.\ Samuelson, A.\ Chedin,
and N.\ Scott 1981. Study of the Ammonia Ice Cloud Layer in the Equatorial
Region of Jupiter from the Infrared Interferometric Experiment on Voyager.
{\it Icarus} {\bf 46}, 233--248.\\

\noindent Moreno, R.\ 1998. {\it Observations millim\'etriques et
submillim\'etriques des plan\`etes g\'eantes. \'Etude de Jupiter apr\`es la
chute de la com\`ete SL9}. Th\`ese de Doctorat, Universit\'e Paris VI, Paris.\\

\noindent Ortiz, J.\ L., G.\ S.\ Orton,
A.\ J.\ Friedson, S.\ T.\ Stewart, B.\ M.\ Fisher, and J.\ R.\ Spencer 1998.
Evolution and persistence of 5-$\mu$m hot spots at the Galileo probe entry
latitude. {\it J.\ Geophys.\ Res.}\ {\bf 103}, 23051--23069.\\

\noindent Owen, T., S.\ Atreya, P.\ Mahaffy, H.\ Niemann, and M.\ H.\ Wong
1997. On the origin of Jupiter's atmosphere and the volatiles on the Medicean
stars. In {\it Three Galileos: The Man, The Spacecraft, The Telescope},
(C.\ Barbieri, J.\ Rahe, T.\ Johnson, and A.\ Sohus, Eds.), pp.~289--297.
Kluwer Academic, Norwell.\\

\noindent Pillinger, C.\ T.\ 1984. Light element stable isotopes in
meteorites---from grams to picograms. {\it Geochim.\ Cosmochim.\ Acta}
{\bf 48}, 2739--2766.\\

\noindent Ragent, B., D.\ S.\ Colburn, K.\ A.\ Rages, T.\ C.\ D.\ Knight,
P.\ Avrin, G.\ S.\ Orton, P.\ A.\ Yanamandra-Fisher, and G.\ W.\ Grams 1998.
The clouds of Jupiter: results of the Galileo Jupiter mission probe
nephelometer experiment. {\it J.\ Geophys.\ Res.}\ {\bf 103}, 22891--22909.\\

\noindent Rinsland, C.\ P., and 11 colleagues 1984. Simultaneous stratospheric
measurements of H$_2$O, HDO, and CH$_4$ from balloon-borne and aircraft
infrared solar absorption spectra and tunable diode laser laboratory spectra of
HDO. {\it J.\ Geophys.\ Res.}\ {\bf 89}, 7259--7266.\\

\noindent Roos, M., and 12 colleagues 1993. The upper clouds of Venus:
Determination of the scale height from NIMS-Galileo infrared data.
{\it Planet.\ Space Sci.}\ {\bf 41}, 505--514.\\

\noindent Roos-Serote, M., and 12 colleagues 1998. Analysis of Jupiter NEB hot
spots in the 4--5\,\micron\ range from Galileo/NIMS observations: measurements
of cloud opacity, water, and ammonia. {\it J.\ Geophys.\ Res.}, {\bf 103},
23023--23041.\\

\noindent Roos-Serote, M., P.\ Drossart, T.\ Encrenaz, R.\ W.\ Carlson, and
L.\ Leader 1999. Constraints on the tropospheric cloud structure of Jupiter
from spectroscopy in the 5-\micron\ region; a comparison between Voyager/IRIS,
Galileo/NIMS and ISO/SWS spectra. {\it Icarus}, in press.\\

\noindent Schaeidt, S.\ G., and 31 colleagues 1996. The photometric calibration
of the ISO Short Wavelength Spectrometer. {\it Astron.\ Astrophys.}\ {\bf 315},
L55--L59.\\

\noindent Scmitt, B., E.\ Quirico,
F.\ Trotta, and W.\ M.\ Grundy 1998. Optical properties of ices from UV to
infrared. In {\it Solar system ices}, (B.\ Shmitt, C.\ de Bergh, and
M.\ Festou, Eds.), pp.\ 199--240. Kluwer Academic Publishers, Dordrecht.\\

\noindent Seiff, A., D.\ B.\ Kirk, T.\ C.\ D.\ Knight, R.\ E.\ Young,
J.\ D.\ Mihalov, L.\ A.\ Young, F.\ S.\ Milos, G.\ Schubert, R.\ C.\ Blanchard,
and D.\ Atkinson 1998. Thermal structure of Jupiter's atmosphere near the edge
of a 5-$\mu$m hot spot in the north equatorial belt. {\it J.\ Geophys.\ Res.}
{\bf 103}, 22857--22889.\\

\noindent Sromovsky, L.\ A., A.\ D.\ Collard, P.\ M.\ Fry, G.\ S.\ Orton,
M.\ T.\ Lemmon, M.\ G.\ Tomasko, and R.\ S.\ Freedman 1998. Galileo probe
measurements of thermal and solar radiation fluxes in the jovian atmosphere.
{\it J.\ Geophys.\ Res.}\ {\bf 103}, 22929--22977.\\

\noindent Terrile, R.\ J., R.\ W.\ Capps, D.\ E.\ Backman, E.\ E.\ Becklin,
D.\ P.\ Cruikshank, C.\ A.\ Beichman, R.\ H.\ Brown, and J.\ A.\ Westfall 1979.
Infrared images of Jupiter at 5-micrometer wavelength during the Voyager 1
encounter. {\it Science} {\bf 204}, 1007--1008.\\

\noindent Thiemens, M.\ H., and R.\ N.\ Clayton 1981. Nitrogen isotopes in the
Allende meteorite. {\it Earth Planet.\ Sci.\ Lett.}\ {\bf 55}, 363--369.\\

\noindent Thode, H.\ G.\ 1940. The vapor pressures, heats of vaporization and
melting points of N$^{14}$ and N$^{15}$ ammonias. {\it J.\ Amer.\ Chem.\ Soc.}
{\bf 62}, 581--583.\\

\noindent Tokunaga, A.\ T, R.\ F.\ Knacke, and S.\ T.\ Ridgway 1980. High
spatial and spectral resolution 10-$\mu$m observations of Jupiter. {\it Icarus}
{\bf 44}, 93--101.\\

\noindent Valentijn E.\ A., and 23 colleagues 1996. The wavelength calibration
and resolution of the SWS. {\it Astron.\ Astrophys.}\ {\bf 315}, L60--L64.\\

%%%%%%%%%%%%%%%%%%%%%%%%%%%%%%%%%%%%%%%%%%%%%%%%%%%%%%%%%%%%%%%%%%%%%%%%%%%%%
\newpage
\noindent {\bf Table I}: Mixing ratios, spectroscopic data and references\\[1 cm]
\begin{tabular}{cccc} 
Gas & Mole Fraction & Line parameters & Broadening coefficient\\ \hline
CH$_4$ & $2.1\times10^{-4}$ (Niemann {\em et al.}\ 1998) & Husson {\em et al.}\ (1992) & Margolis (1996)\\
& & Lellouch {\em et al.}\ (1987) & \\
NH$_3$ & to be determined & and & Brown and Peterson (1994)\\
 & & Husson {\em et al.}\ (1992) & \\
PH$_3$ & $7\times10^{-7}$ (Kunde et {\em at al.}\ 1982) & Husson {\em et al.}\ (1992)& Levy {\em et al.}\ (1993)\\
CH$_3$D & $2.5\times10^{-7}$ (Encrenaz {\em et al.}\ 1996) & Husson {\em et al.}\ (1992) & same as CH$_4$\\
H$_2$O & to be determined & Husson {\em et al.}\ (1992) & Brown and Plymate (1996)\\
GeH$_4$ & $3.0\times10^{-10}$ (Drossart {\em et al.}\ 1996) & Husson {\em et al.}\ (1992) & same as CH$_4$\\
AsH$_3$ & $2.5\times10^{-10}$ (Noll {\em et al.}\ 1990) & Dana {\em et al.}\ (1993) & same as PH$_3$\\ \hline
\end{tabular}

\newpage
\noindent {\bf Table II}: \ratio\ ratios and fractionation coefficients for three different temperature profiles\\[1 cm]
\begin{tabular}{cccccccc}
Lineshape factor & cloud transmission & $\delta$ (\%) & $\alpha$ & $\delta$ (\%) & $\alpha$ & $\delta$ (\%) & $\alpha$\\
\multicolumn{2}{c}{ } & \multicolumn{2}{c}{Cold profile} & \multicolumn{2}{c}{Nominal profile} & \multicolumn{2}{c}{Hot profile}\\ \hline 
 & 1.0 & -41 & 1.17 & -49 & 1.23 & -73 & 1.44\\[-0.2cm]
cutoff & 0.9 & -39 & 1.15 & -48 & 1.22 & -73 & 1.44\\[-0.2cm]
 at & 0.8 & -36 & 1.13 & -47 & 1.21 & -73 & 1.43\\[-0.2cm]
 50 \cm\ & 0.7 & -32 & 1.11 & -46 & 1.20 & -72 & 1.43\\[-0.2cm]
 & 0.6 & -25 & 1.08 & -45 & 1.19 & -72 & 1.42\\ \hline
 & 1.0 & -45 & 1.20 & -52 & 1.25 & -75 & 1.45\\[-0.2cm]
cutoff & 0.9 & -43 & 1.18 & -52 & 1.25 & -74 & 1.45\\[-0.2cm]
 at & 0.8 & -40 & 1.16 & -50 & 1.23 & -74 & 1.44\\[-0.2cm]
 20 \cm\ & 0.7 & -36 & 1.13 & -49 & 1.22 & -73 & 1.44\\[-0.2cm]
 & 0.6 & -29 & 1.10 & -48 & 1.21 & -73 & 1.43\\ \hline
 & 1.0 & -40 & 1.17 & -50 &  1.24 & -72 & 1.43\\[-0.2cm]
 & 0.9 & -39 & 1.15 & -49 & 1.22 & -72 & 1.43\\[-0.2cm]
$\chi(\nu)=e^{-(\nu-\nu_0)/80}$ & 0.8 & -36 & 1.14 & -47 & 1.21 & -72 & 1.42\\[-0.2cm]
 & 0.7 & -31 & 1.11 & -47 & 1.21 & -71 & 1.42\\[-0.2cm]
 & 0.6 & -25 & 1.08 & -46 & 1.20 & -71 & 1.41\\ \hline
 & 1.0 & -40 & 1.17 & -50 & 1.25 & -72 & 1.43 \\[-0.2cm]
 & 0.9 & -38 & 1.16 & -49 & 1.24 & -71 & 1.43\\[-0.2cm]
$\chi(\nu)=e^{-(\nu-\nu_0)/30}$ & 0.8 & -36 & 1.14 & -48 & 1.23 & -71 & 1.42\\[-0.2cm]
 & 0.7 & -31 & 1.11 & -46 & 1.21 & -71 & 1.42\\[-0.2cm]
 & 0.6 & -25 & 1.09 & -45 & 1.20 & -71 & 1.41\\ \hline
\end{tabular}

\newpage
\section*{Figure Captions}

\noindent {\bf Figure 1} Normalized contribution functions for the \nh\ Q-branch center at 10.73\,\micron\ (solid line), a \nh\ line at 10.83\,\micron\ (dashed line) and in the ``continuum'' at 10.91\,\micron\ (dash-dotted line). These functions indicate the pressure levels probed by the different \nh\ lines.\\

\noindent {\bf Figure 2} Panel a: Comparison between observed (solid line) and synthetic spectra calculated for 10-\micron sized \nh-ice particles (dashed line) and for grey particles (dotted line) in the 10-\micron\ range. Panel b: Absorption cross section of \nh-ice spheres of 10-\micron\ radius (solid line) and 100-\micron\ radius (dashed line).\\

\noindent {\bf Figure 3} \nh\ distributions derived from the 10-\micron\ region for three different temperature profiles: the nominal Galileo-based profile (solid line), a hot profile (nominal profile $+2\,$K) (dotted line), a cold profile (nominal profile $-2\,$K) (dashed-dotted line).\\

\noindent {\bf Figure 4} Respective influence of the absorptions due to ammonia and water on the 5-\micron\ window spectrum along with the contributions functions. Panels a and c: ISO-SWS spectrum (solid line) compared with synthetic spectra (dashed-dotted line) in which are included only the absorptions of water (a) and ammonia (c). Panel b: contribution functions in the line of water at 5.21\,\micron\ (dash-dotted line) and 4.84\,\micron\ (solid line). Panel d: contribution functions in the ammonia line at 5.32\,\micron\ (dash-dotted line), 5.2\,\micron\ (dashed line) and 4.86\,\micron (solid line).\\

\noindent {\bf Figure 5} Comparison between the observed spectrum (solid lines) and synthetic spectra calculated for an infinitely thin cloud layer located at 1\,bar with a transmission of 8\%. In the first model (dotted line) the ammonia vertical distribution is contrained to fit the ammonia lines at 5.32, 5.20 and shortward of 5\,\micron. In the second model (dashed line) the ammonia vertical distribution is constrained to fit the observations longward of 5.25\,\micron. The arrows indicate \nh-lines.\\

\noindent {\bf Figure 6} Comparison between observed (solid lines) and synthetic spectra calculated for a vertically uniform ammonia profile (dotted lines) and an ammonia profile decreasing with altitude (dashed line). The synthetic spectra are calculated for an infinitely thin cloud based at 1\,bar and transmission 8\%. The arrows indicate \nh-lines.\\

\noindent {\bf Figure 7} Same as Fig.~7 for a cloud layer located at 1.9\,bar with scale height ${\rm H_p=0.6\times H_g}$ and transmission 8\%. The arrows indicate \nh-lines.\\

\noindent {\bf Figure 8} Best fit \nh\ distributions derived from the 5-\micron\ window for the spatially homogeneous model (solid line) and for the hot spots region in the spatially heterogeneous model (dashed line). The results of Folkner {\em et al.}\ (1998) are displayed in solid triangles, anf the results of Sromovsky {\em et al.}\ (1998) in open circles.\\

\noindent {\bf Figure 9} Comparison of calculations with empirically accounting (solid line) and without accounting for multiple scattering. The atmospheric model comes from category 8 spectra of Carlson {\em et al.}\ (1993). The calculations are convolved to the IRIS resolution.\\

\noindent {\bf Figure 10} Comparison between the observed spectrum (solid lines) and the synthetic spectrum (dashed lines) calculated with the spatially heterogeneous model described in section 3.1.2, for an infinitely thin cloud layer located at 1\,bar (panel a) and for a cloud layer located at 1.9\,bar with scale height ${\rm H_p=0.4\times H_g}$.\\

\noindent {\bf Figure 11} Test for the Galileo probe derived vertical profiles of \nh\ (dotted line Sromovsky {\em et al.}\ (1998); dashed line Folkner {\em et al.}\ (1998)) against the ISO-SWS data (solid line).\\

\noindent {\bf Figure 12} Comparison between the observed spectrum at 4.8--5.5\,\micron\ (solid lines) and a synthetic spectrum (dashed lines) calculated with the thick water cloud located at 5\,bar. For \nh\ the spatially heterogenous model is used.\\

\noindent {\bf Figure 13} Comparison at 9.5--11.5\,\micron\ between the observed spectrum (solid line) and synthetic spectra calculated for $\delta=0\%$ (dotted lines) and $\delta=-50\%$ (dashed line). The arrows indicate \inh-lines.\\

\noindent {\bf Figure 14} Zoom on the 10.88- and 11.07-\micron\ \rnh\ and \inh\ lines. The observed spectrum (solid line) is compared with a synthetic spectrum calculated for a terrestrial isotopic ratio (dashed line) and with a synthetic spectrum calculated for $\delta=-50\%$ (dashed-dotted line).\\

\noindent {\bf Figure 15} Comparison at 9.5--11.5\,\micron\ between observed (solid line) and synthetic spectra (dotted line) calculated for temperature profile 8\,K colder than the nominal Galileo-based temperature profile. The ammonia vertical profile is constrained to fit the \rnh\ at 10.88\,\micron\ and the surrounding continuum. The \ratio\ isotopic ratio is terrestrial.

\newpage
\pagestyle{empty}
\begin{figure}[!h]
  \begin{center}
  \leavevmode
  \centerline{\epsfig{file=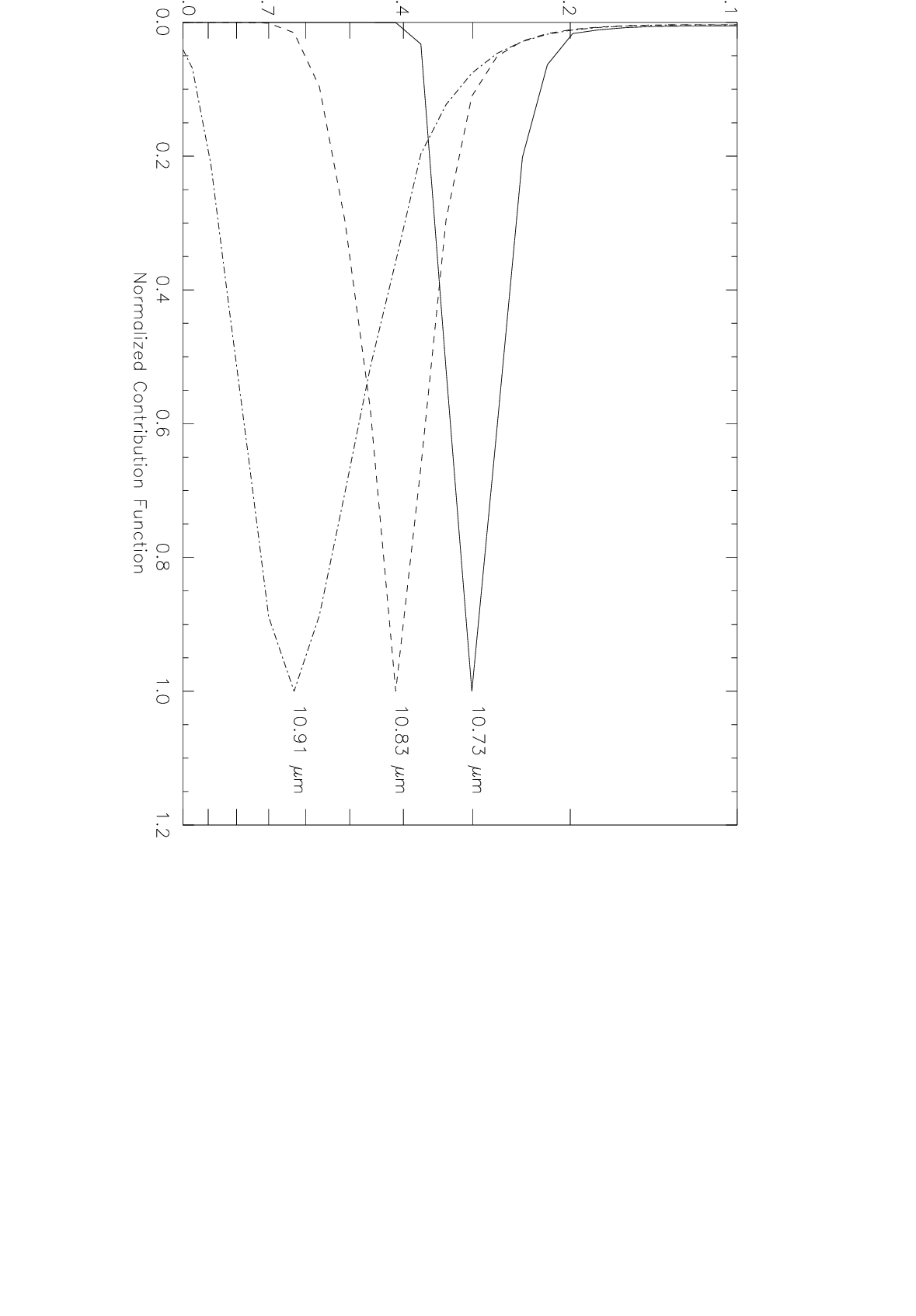}}
  \end{center}
\end{figure}

\newpage
\thispagestyle{empty}
\begin{figure}[!h]
  \begin{center}
  \leavevmode
  \centerline{\epsfig{file=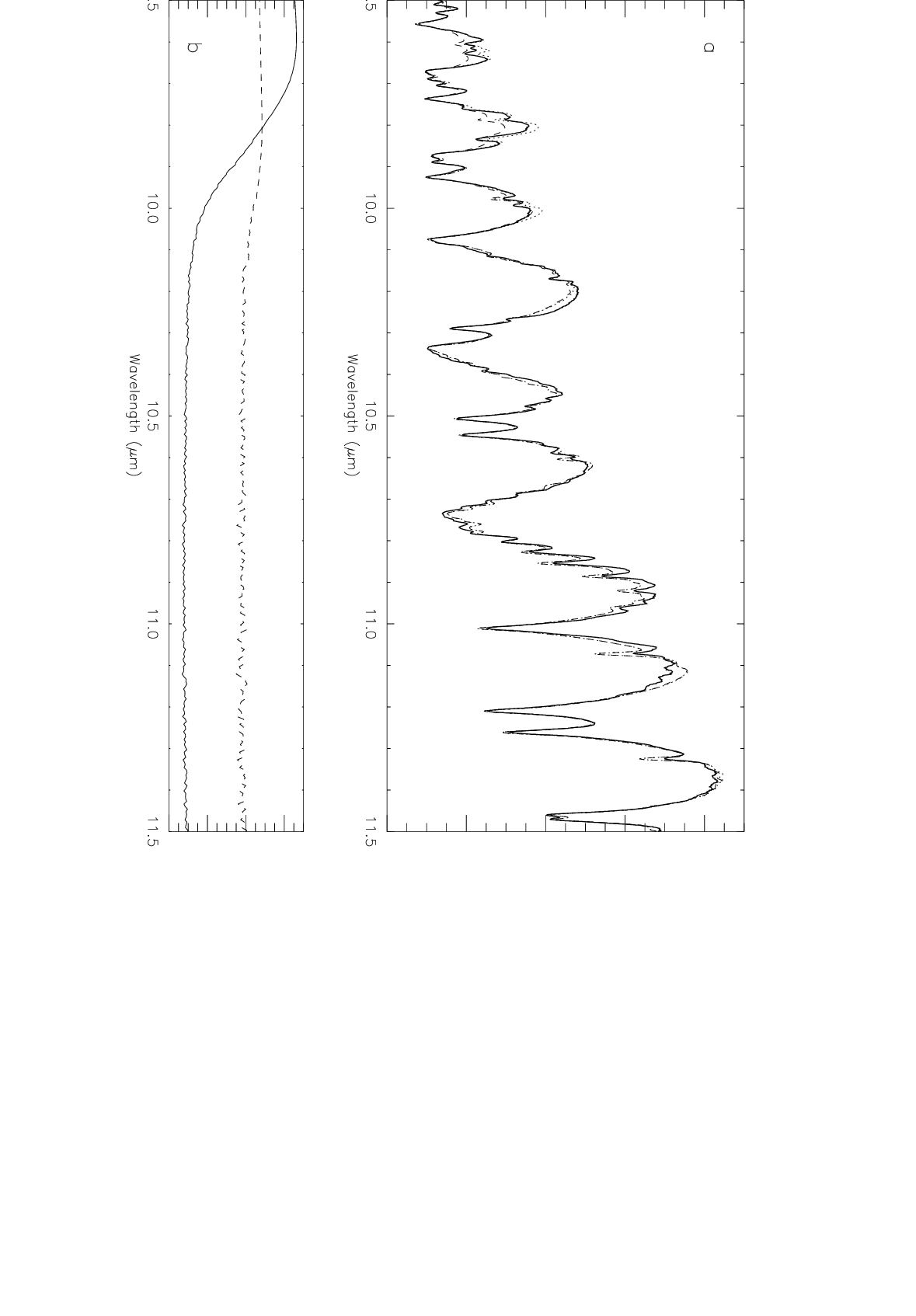}}
  \end{center}
\end{figure}

\newpage
\thispagestyle{empty}
\begin{figure}[!h]
  \begin{center}
  \leavevmode
  \centerline{\epsfig{file=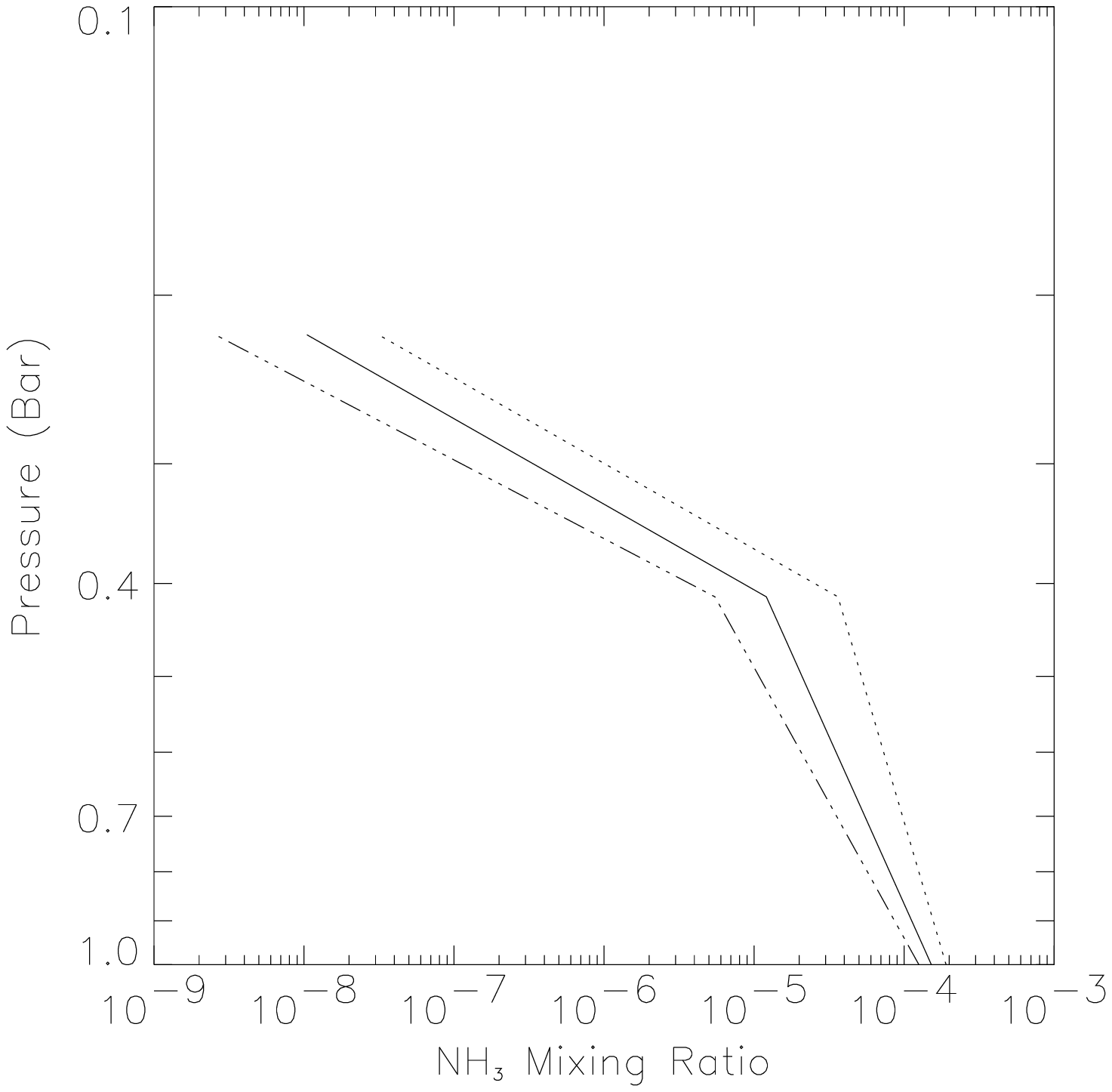}}
  \end{center}
\end{figure}

\newpage
\thispagestyle{empty}
\begin{figure}[!h]
  \begin{center}
  \leavevmode
  \centerline{\epsfig{file=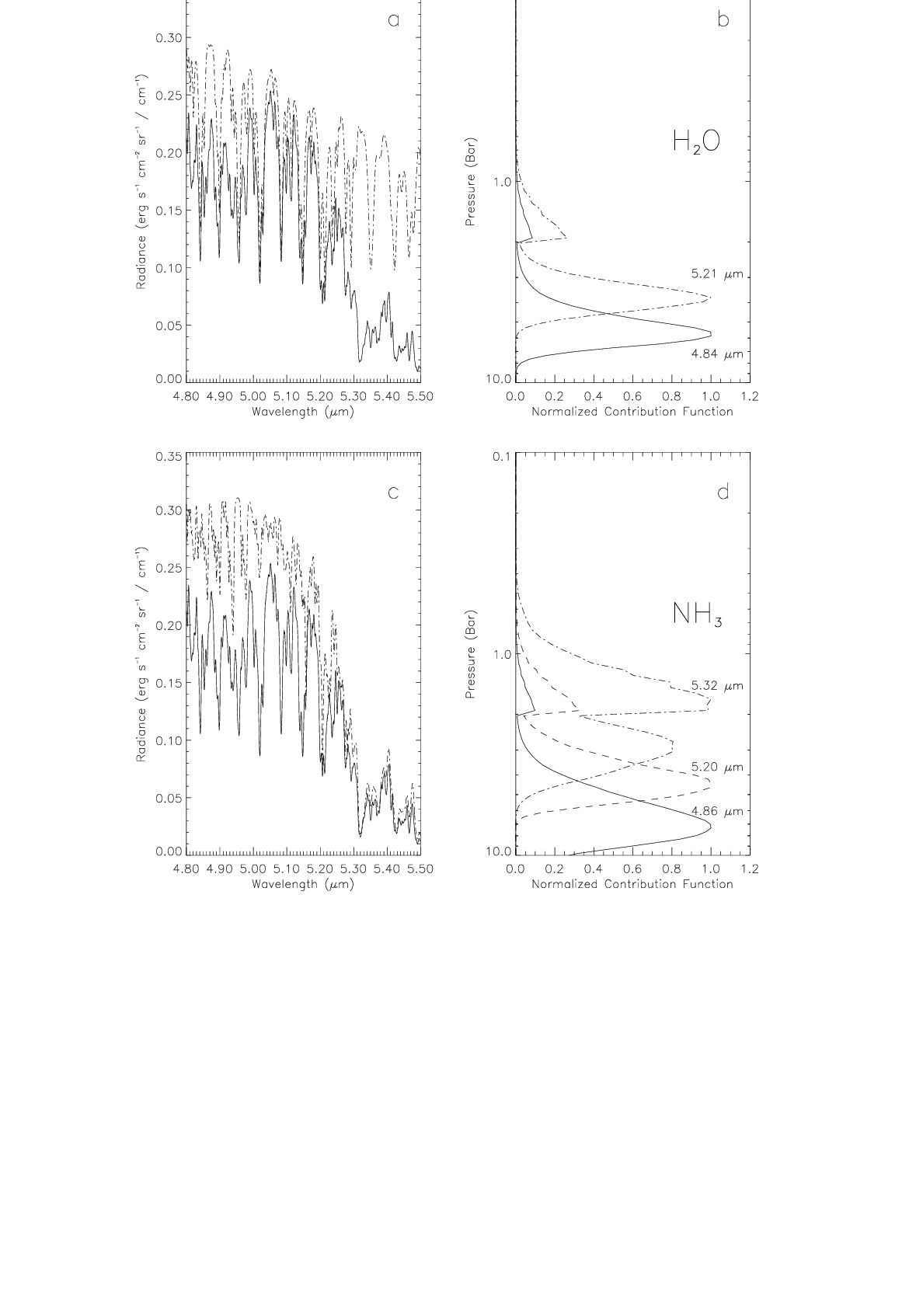}}
  \end{center}
\end{figure}

\newpage
\thispagestyle{empty}
\begin{figure}[!h]
  \begin{center}
  \leavevmode
  \centerline{\epsfig{file=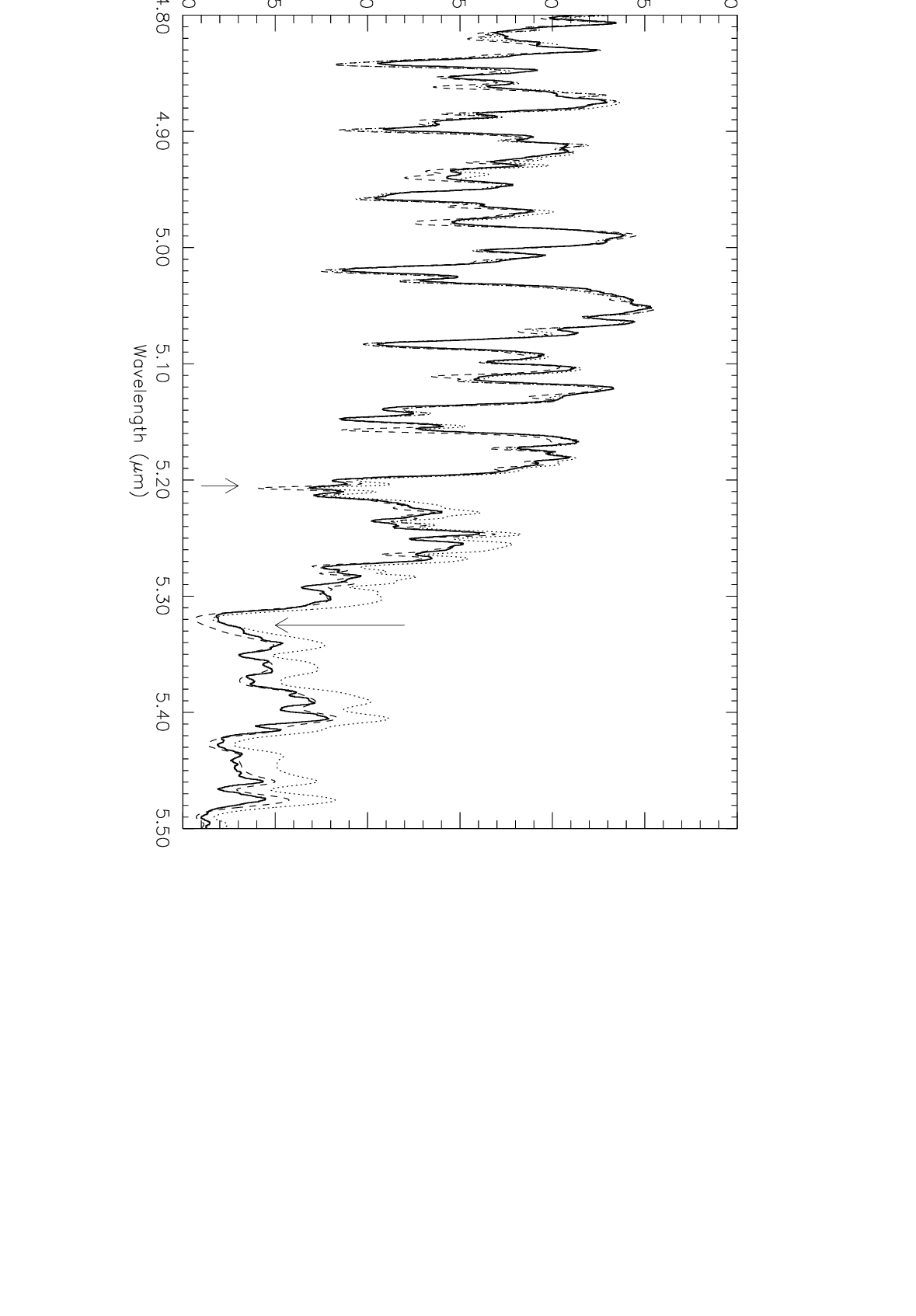}}
  \end{center}
\end{figure}

\newpage
\thispagestyle{empty}
\begin{figure}[!h]
  \begin{center}
  \leavevmode
  \centerline{\epsfig{file=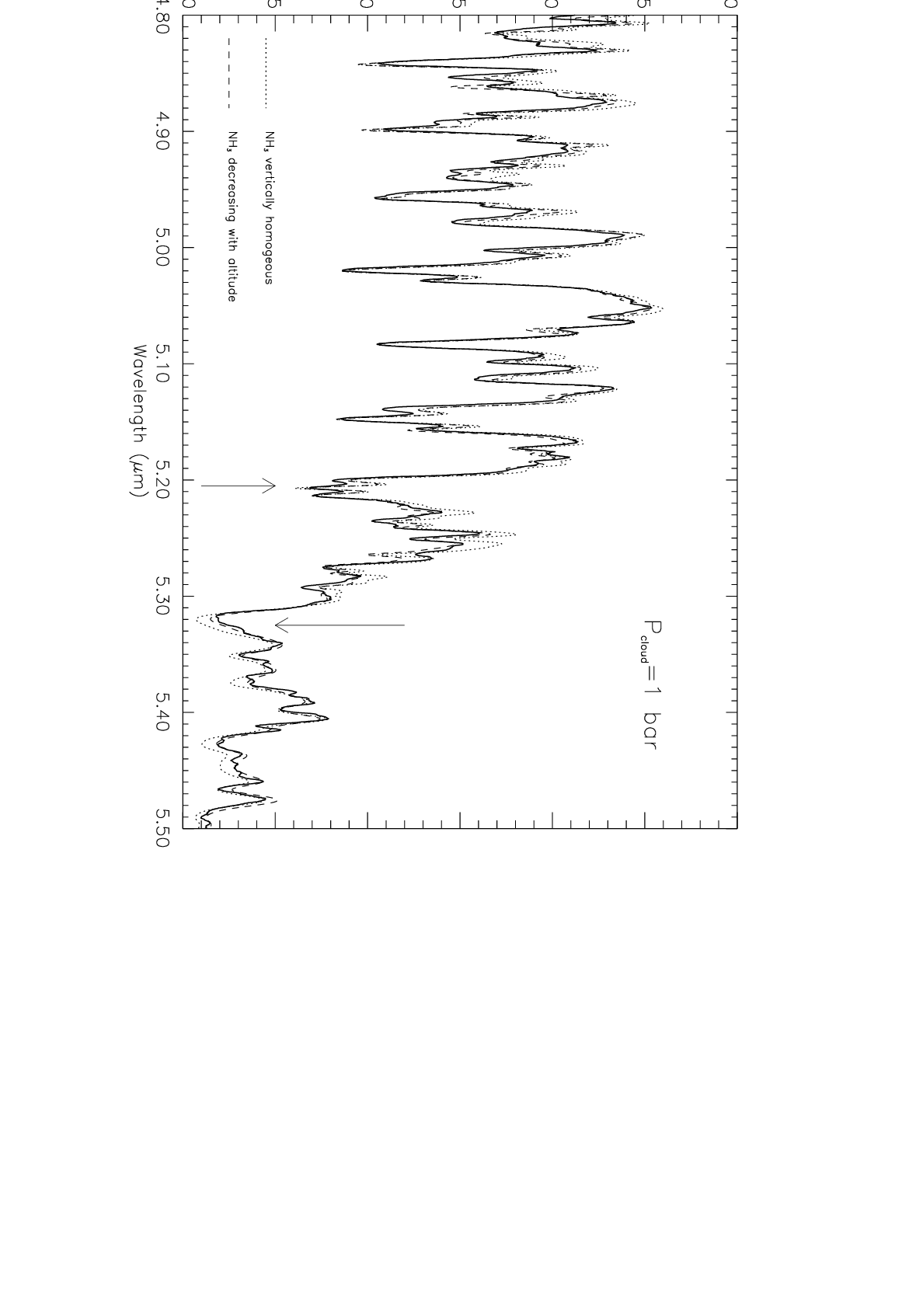}}
  \end{center}
\end{figure}

\newpage
\thispagestyle{empty}
\begin{figure}[!h]
  \begin{center}
  \leavevmode
  \centerline{\epsfig{file=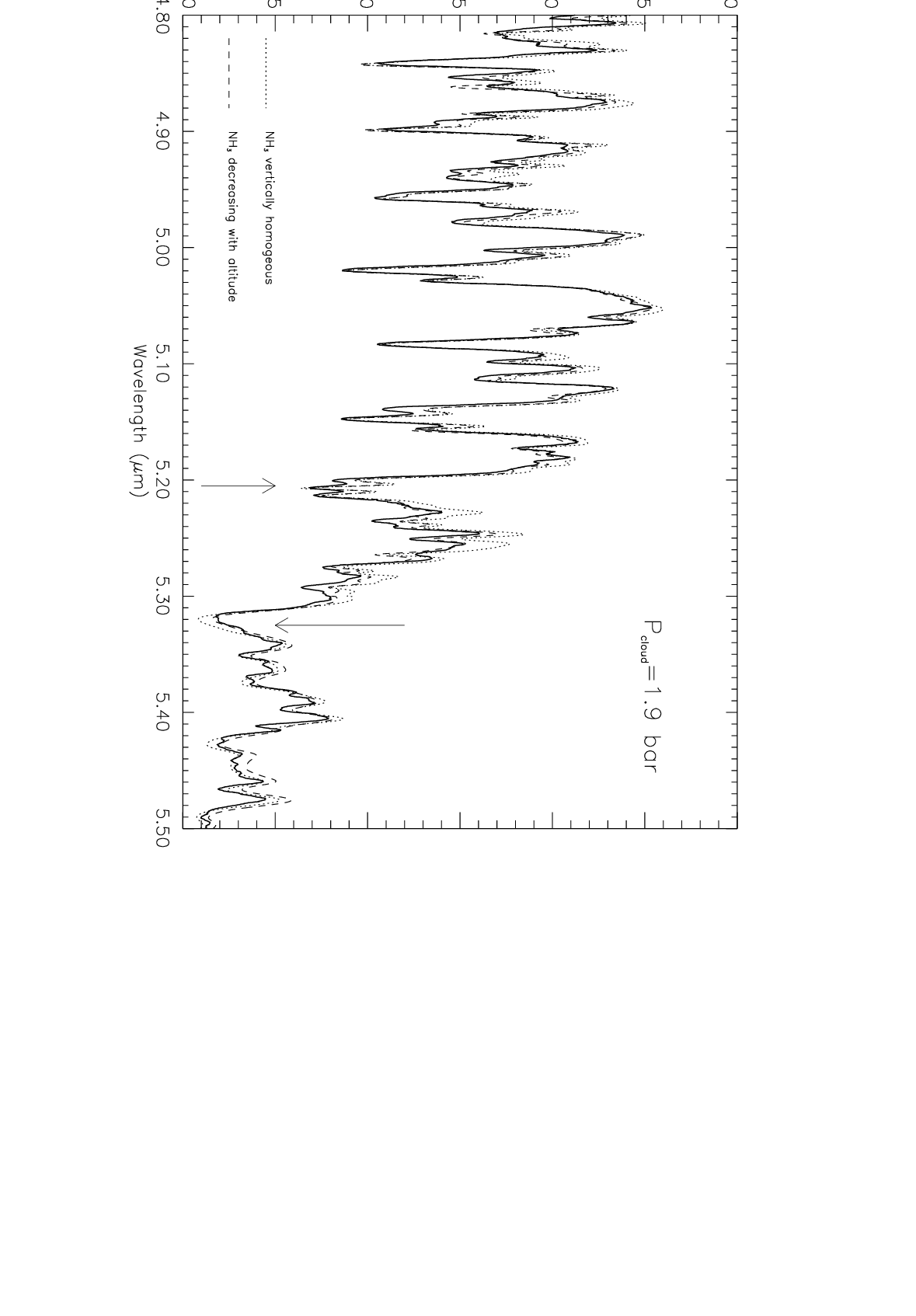}}
  \end{center}
\end{figure}

\newpage
\thispagestyle{empty}
\begin{figure}[!h]
  \begin{center}
  \leavevmode
  \centerline{\epsfig{file=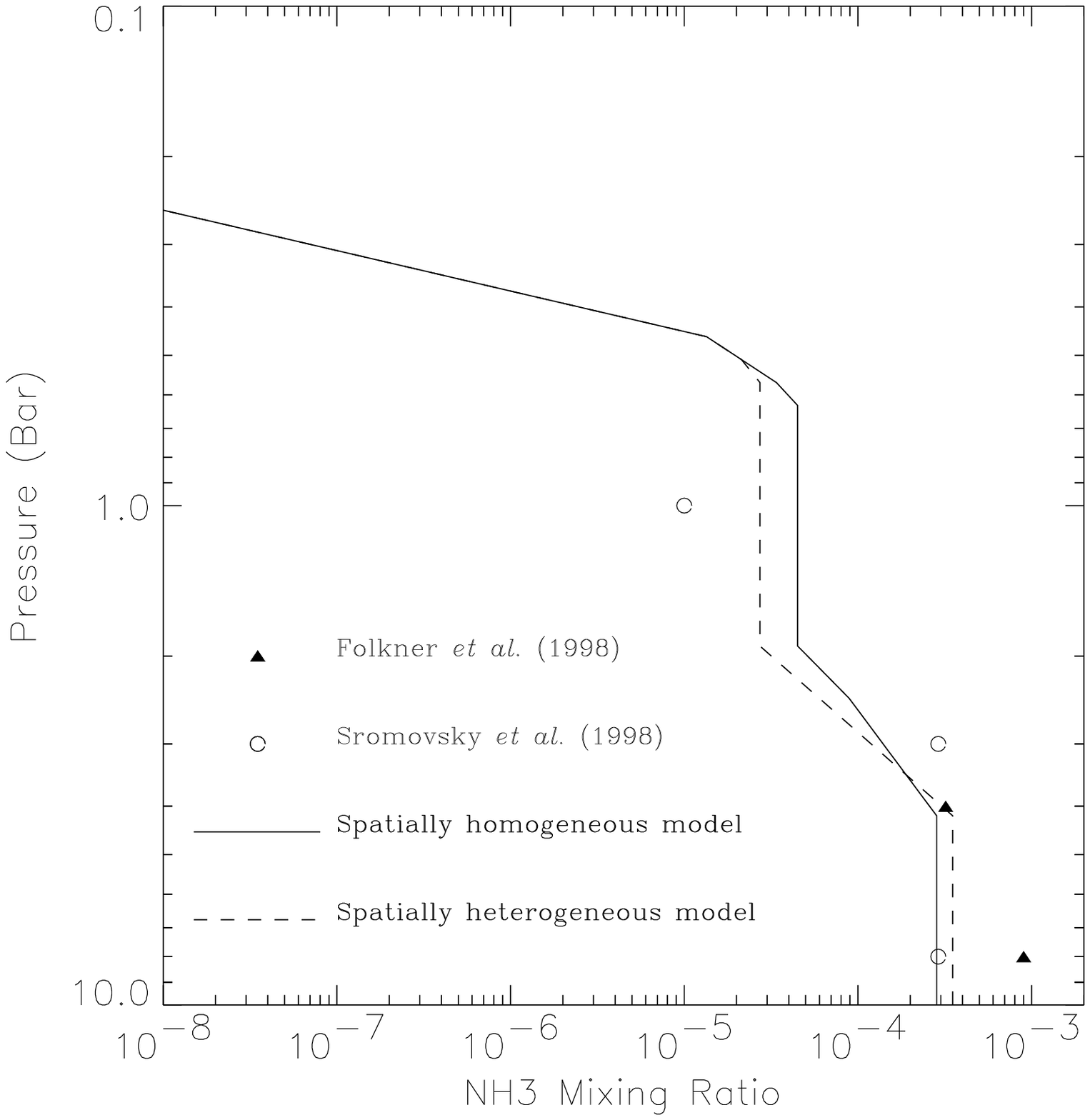}}
  \end{center}
\end{figure}

\newpage
\thispagestyle{empty}
\begin{figure}[!h]
  \begin{center}
  \leavevmode
  \centerline{\epsfig{file=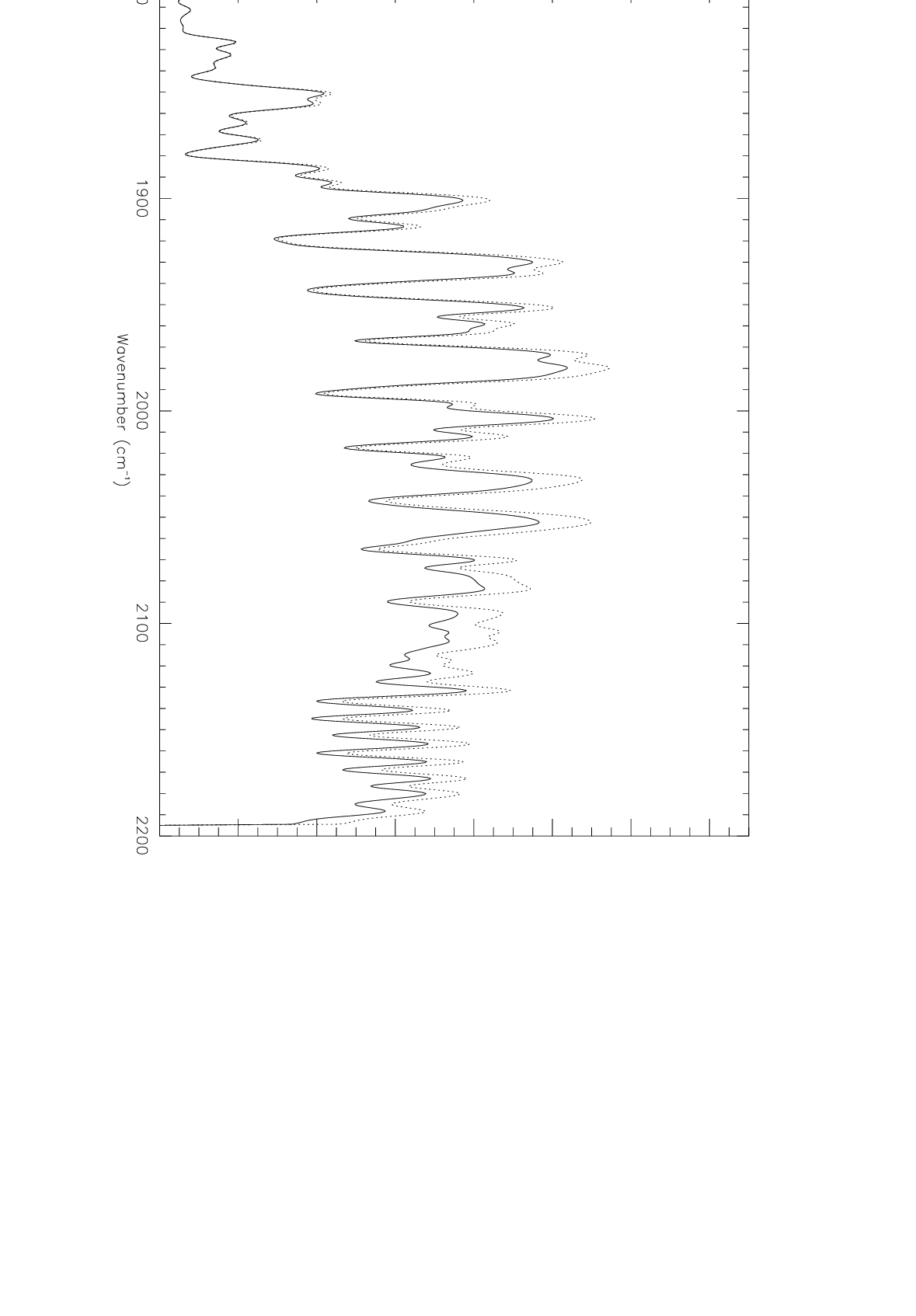}}
  \end{center}
\end{figure}

\newpage
\thispagestyle{empty}
\begin{figure}[!h]
  \begin{center}
  \leavevmode
  \centerline{\epsfig{file=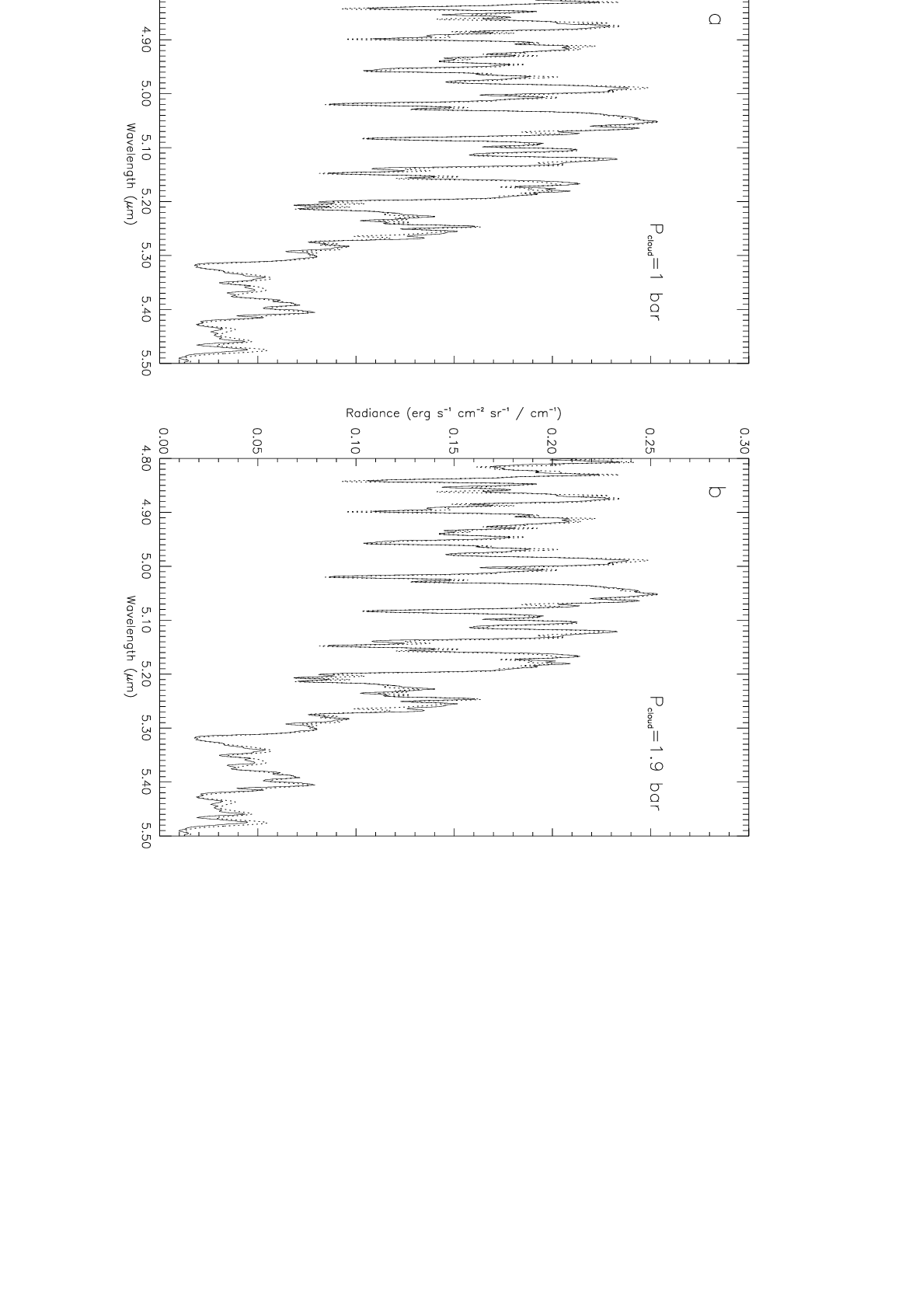}}
  \end{center}
\end{figure}

\newpage
\thispagestyle{empty}
\begin{figure}[!h]
  \begin{center}
  \leavevmode
  \centerline{\epsfig{file=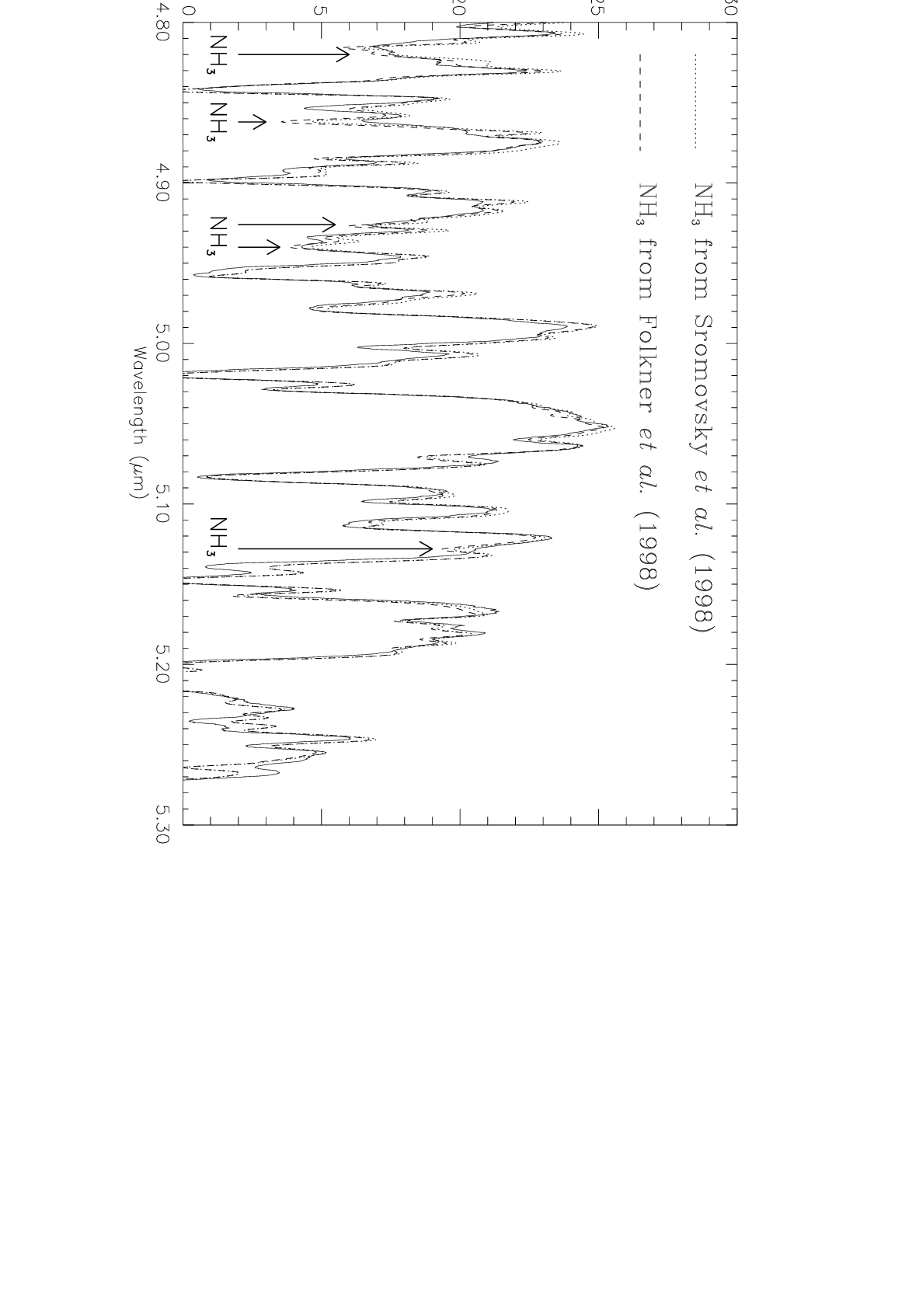}}
  \end{center}
\end{figure}

\newpage
\thispagestyle{empty}
\begin{figure}[!h]
  \begin{center}
  \leavevmode
  \centerline{\epsfig{file=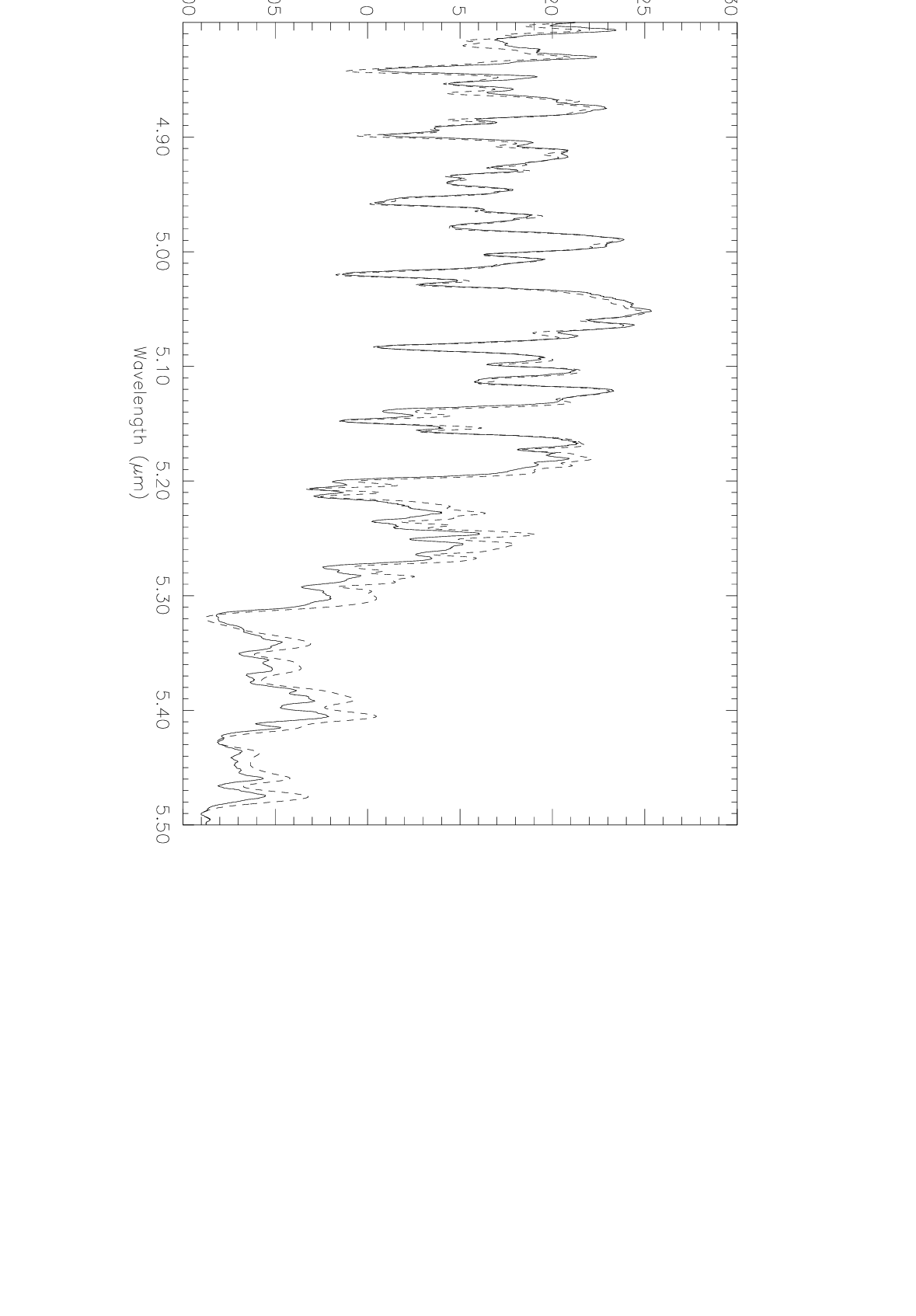}}
  \end{center}
\end{figure}

\newpage
\thispagestyle{empty}
\begin{figure}[!h]
  \begin{center}
  \leavevmode
  \centerline{\epsfig{file=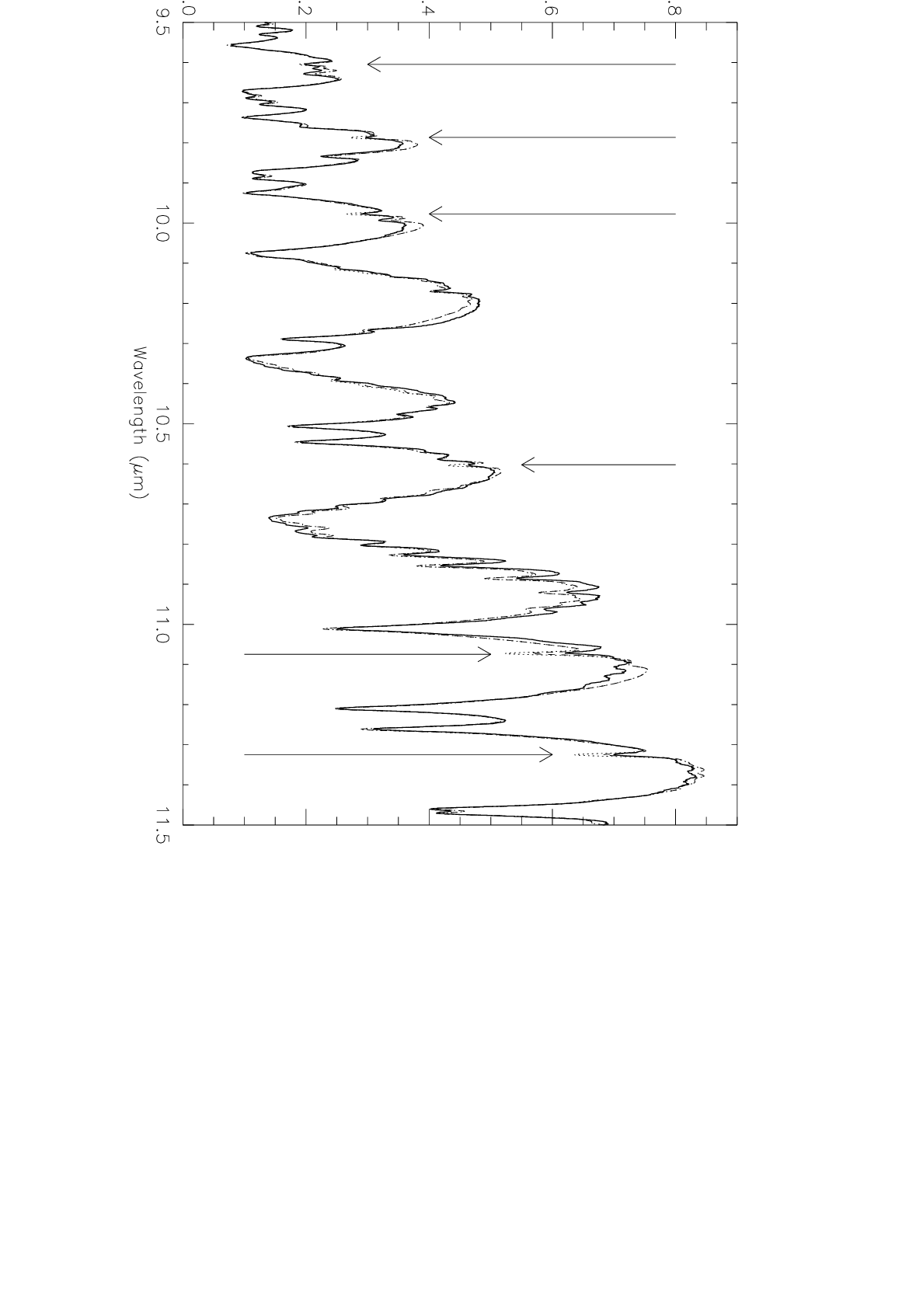}}
  \end{center}
\end{figure}

\newpage
\thispagestyle{empty}
\begin{figure}[!h]
  \begin{center}
  \leavevmode
  \centerline{\epsfig{file=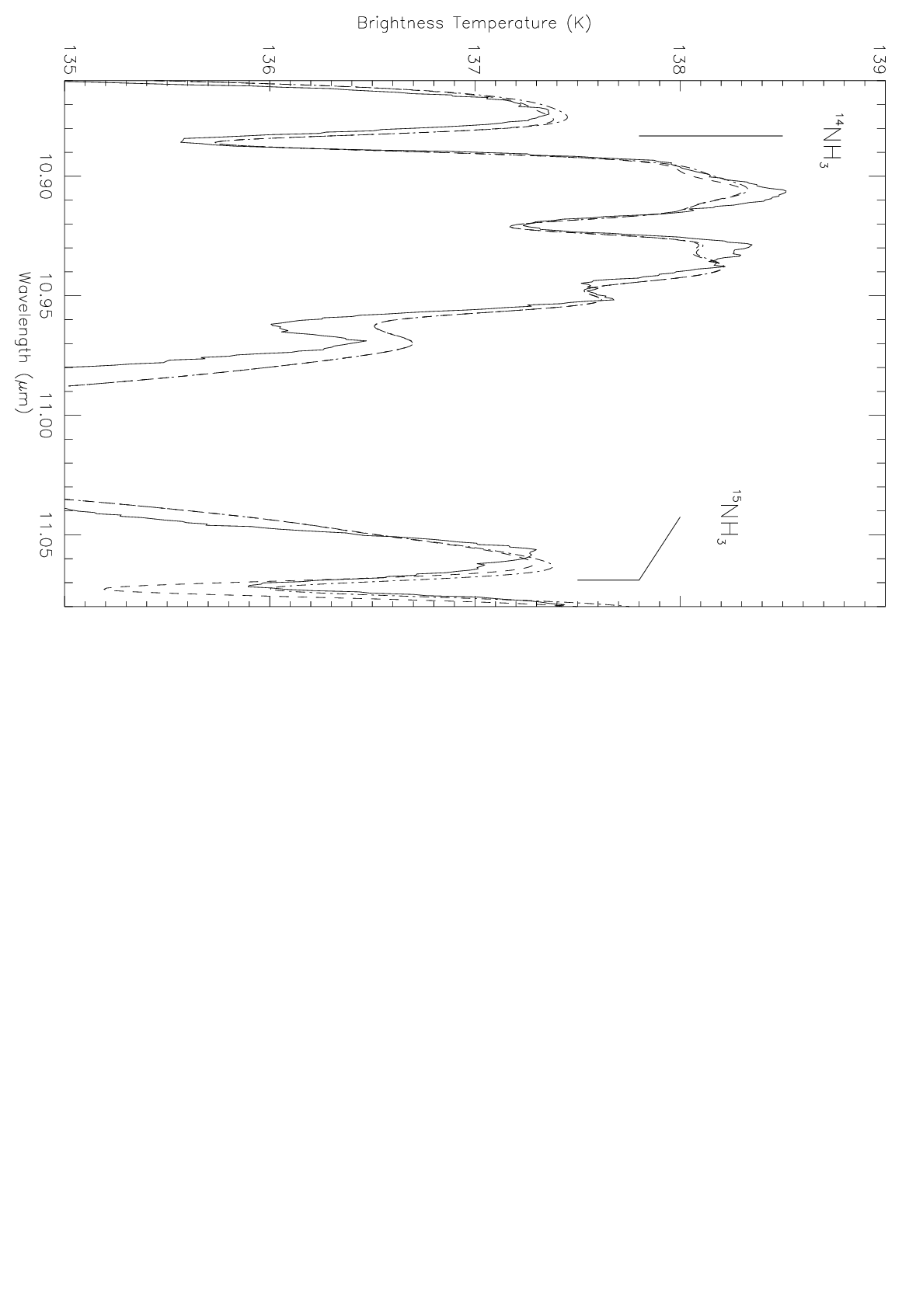}}
  \end{center}
\end{figure}

\newpage
\thispagestyle{empty}
\begin{figure}[!h]
  \begin{center}
  \leavevmode
  \centerline{\epsfig{file=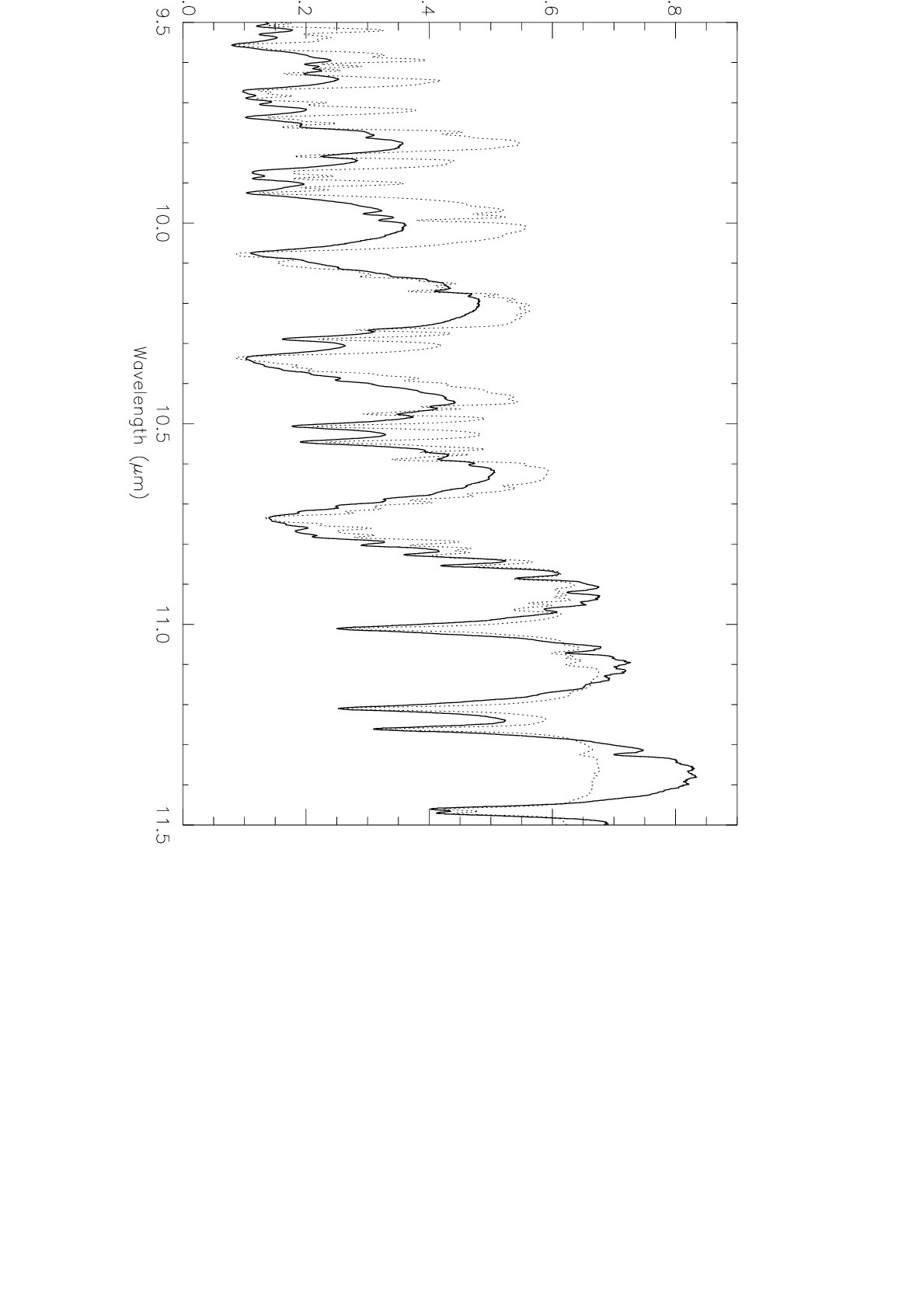}}
  \end{center}
\end{figure}

\end{document}